\documentclass[a4paper,fleqn]{cas-sc}

\usepackage[authoryear,longnamesfirst]{natbib}
\usepackage{lineno}
\usepackage{graphicx,subfigure,float}
\usepackage{makecell}

\def\tsc#1{\csdef{#1}{\textsc{\lowercase{#1}}\xspace}}
\tsc{WGM}
\tsc{QE}
\tsc{EP}
\tsc{PMS}
\tsc{BEC}
\tsc{DE}

\begin{document}
\let\WriteBookmarks\relax
\def\floatpagepagefraction{1}
\def\textpagefraction{.001}
\shorttitle{}
\shortauthors{Haoyan Yang and others}

\title [mode = title]{The Research and Development of New Electronics System and its Testing on the JNE-1ton Prototype Detector }                      

\author[1]{Haoyan Yang}
\author[2,3]{Yuzi Yang}
\cormark[1]
\ead{yangyuzi@lzu.edu.cn}
\author[3]{Yapeng Wang}
\author[3]{Changxu Wei}
\author[3]{Haoyang Fu}
\author[3]{Haozhe Sun}
\author[2]{Juntao Liu}
\author[2]{Zhiyi Liu}
\author[1]{Tao Xue}
\cormark[1]
\ead{xuetaothu@tsinghua.edu.cn}
\author[1]{Jianmin Li}
\author[1]{Yinong Liu}
\author[3]{Zhe Wang}
\author[3]{Shaomin Chen}
\cormark[1]
\ead{chenshaomin@tsinghua.edu.cn}

\affiliation[1]{organization={Key Laboratory of Particle \& Radiation Imaging},
                addressline={Tsinghua University}, 
                city={Beijing},
                postcode={100084}, 
                country={P.R. China}}
\affiliation[2]{organization={School of Nuclear Science and Technology \& MOE Frontiers Science Center for Rare Isotopes},
                addressline={Lanzhou University}, 
                postcode={730000}, 
                city={Lanzhou},
                country={P.R. China}}
\affiliation[3]{organization={Department of Engineering Physics \& Center for High Energy Physics},
                addressline={Tsinghua University}, 
                city={Beijing},
                postcode={100084}, 
                country={P.R. China}}

\cortext[cor1]{Corresponding author}

\begin{abstract}
The Jinping Neutrino Experiment (JNE), a next-generation neutrino observatory under construction at the China Jinping Underground Laboratory II (CJPL-II), requires high-precision waveform-based event reconstruction, imposing stringent demands on its readout electronics. To meet these requirements, we have developed a high-performance readout system featuring 1 GSa/s real-time sampling, 14-bit physical resolution with an effective number of bits (ENOB) of 10.6, a total data throughput of 64 Gbps, and a deterministic zero-delay clock distribution architecture.
The new single-crate 64-channel system (PDS1500) was validated through bench tests and deployment on the upgraded JNE-1ton prototype detector. Its performance was further evaluated against a commercial reference system. The results demonstrate that all key metrics meet the JNE experimental requirements: zero data loss within a 1000 ns acquisition window, baseline noise reduced to one-third of the reference level, timing drift limited to 0.3 ns across power cycles, and an energy threshold as low as 0.1 MeV, enabling the detection of low-energy solar neutrinos. While the 14-bit physical resolution provides significantly higher waveform fidelity, the overall energy resolution in this test remains dominated by the intrinsic limitations of the JNE-1ton detector, as expected.
Furthermore, the modular architecture provides the throughput and scalability required to support the full-scale 3000-channel JNE detector. These results collectively demonstrate that the newly developed electronics system fully satisfies the technical requirements of the future JNE experiment.
\end{abstract}


\maketitle
\section{Introduction}

\begin{figure}[h]
	\centering
	\includegraphics[width=.9\textwidth]{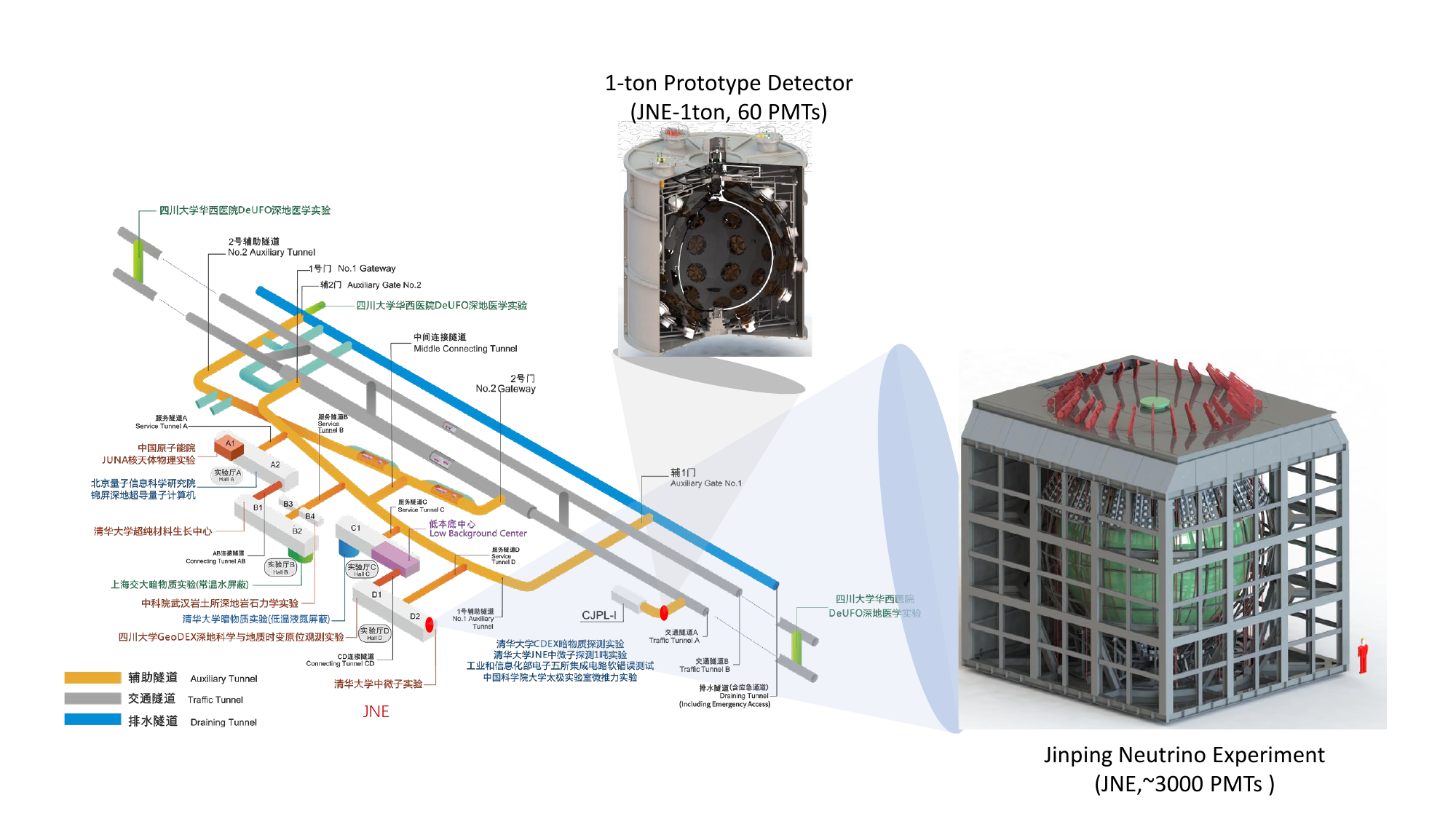}
	\caption{The locations of JNE and JNE-1ton in Jinping and the schematic diagram of their detector.}
	\label{fig:CJPLJNE}
\end{figure}

The Jinping Neutrino Experiment (JNE), a next-generation neutrino observatory under construction~\cite{Jinping:2016iiq}, is located at the China Jinping Underground Laboratory-II (CJPL-II)~\cite{Ma:2021uzi}—the world’s deepest operational underground laboratory ($\sim$2400 m), as shown in Figure~\ref{fig:CJPLJNE}. 
The detector is a 10-meter-diameter acrylic sphere, containing 500 tons of slow liquid scintillator target and equipped with approximately 3000 photomultiplier tubes (PMTs) with light concentrators~\cite{Guo:2017nnr,Ouyang:2025phk}.

By analyzing the digitized waveforms from each PMT, JNE achieves precise measurement of photon count and arrival time per channel~\cite{wu2023performance,Luo:2023reconstruction,WANG2026170986}. This enables high-fidelity separation of Cherenkov radiation from isotropic scintillation light~\cite{SNO:2023cnz,BOREXINO:2021xzc}, thereby facilitating high-resolution reconstruction of neutrino interaction vertex, energy deposition, and directionality. 
These capabilities establish a robust foundation for precision  neutrino physics. JNE will detect solar neutrinos~\cite{Xu:2022wcq,Shao:2022yjc}, their day-night asymmetry~\cite{Super-Kamiokande:2023jbt}, geoneutrinos from U/Th decays in the Earth~\cite{Bellini:2013wsa,wan2017geoneutrinos,wang2020hunting}, supernova and supernova remnant neutrinos~\cite{DeGouvea:2020ang,wei2017discovery}, and search for neutrinoless double-beta decay~\cite{Dolinski:2019nrj,fu2024comparison}.

Unlike the charge-time readout mode that only records the integrated charge quantity of the PMT and the trigger time~\cite{AN2016133,BOGER2000172,Albanese_2021,ALIMONTI2009568}, the waveform acquisition mode needs to simultaneously complete the digital acquisition and real-time processing of the complete analog waveforms of thousands of PMT outputs within a time window of approximately 1000 ns. 
From a physical perspective, the weak neutrino signals must be effectively identified and extracted in the presence of strong natural radioactive background.
For these reasons, the new electronic system must possess high analog bandwidth, high trigger rate, low processing delay and excellent continuous data throughput capability. Therefore, this electronic readout system faces severe technical challenges.

To meet the requirements of JNE, a new generation of electronic systems has been developed. 
The newly developed electronic single-chassis integrates a 64-channel high-speed and high-precision waveform digitization and high-bandwidth data readout system, consisting of 8 waveform digitization boards, 1 clock and trigger distribution board, and 1 PCIe data acquisition board. 
Each waveform digitization board integrates 8 analog sampling channels, supporting a 1 GSa/s sampling rate and 14-bit resolution.
The clock and trigger distribution board provides high-precision synchronous clock signals for the system, with a phase deviation of less than 200 ps between channels at 1 GHz and a clock jitter of less than 10 ps, ensuring the time consistency of multi-channel sampling. 
The PCIe acquisition board supports an 8-lane PCIe Gen3 interface, with a maximum data transmission bandwidth of 64 Gbps, capable of supporting continuous data transmission requirements at a maximum trigger frequency of up to 50 kHz.

This article provides a detailed account of the research and testing process of the electronic system, 
to ensure that it meets the technical requirements of the JNE. To conduct a more comprehensive assessment of the system's performance, the single-crate 64-channel module (PDS1500) of this new electronic system is applied to the upgraded 1-ton prototype detector of JNE (JNE-1ton, Figure~\ref{fig:CJPLJNE}). 
The comprehensive performance indicates that the new system has met the technical requirements of the future JNE experiment.

The structure of the article is as follows.
Section 2 introduces the design of the new electronic system; Section 3 presents the testing procedure of this system in a one-ton prototype detector, and conducts a comparative analysis with the CAEN system V1751; Section 4 looks forward to the design of the system in the future JNE; Section 5 is the summary.

\section{Electronics system}

\subsection{Hardware architecture}
\begin{figure}[h]
    \centering
    \includegraphics[width=0.9\linewidth]{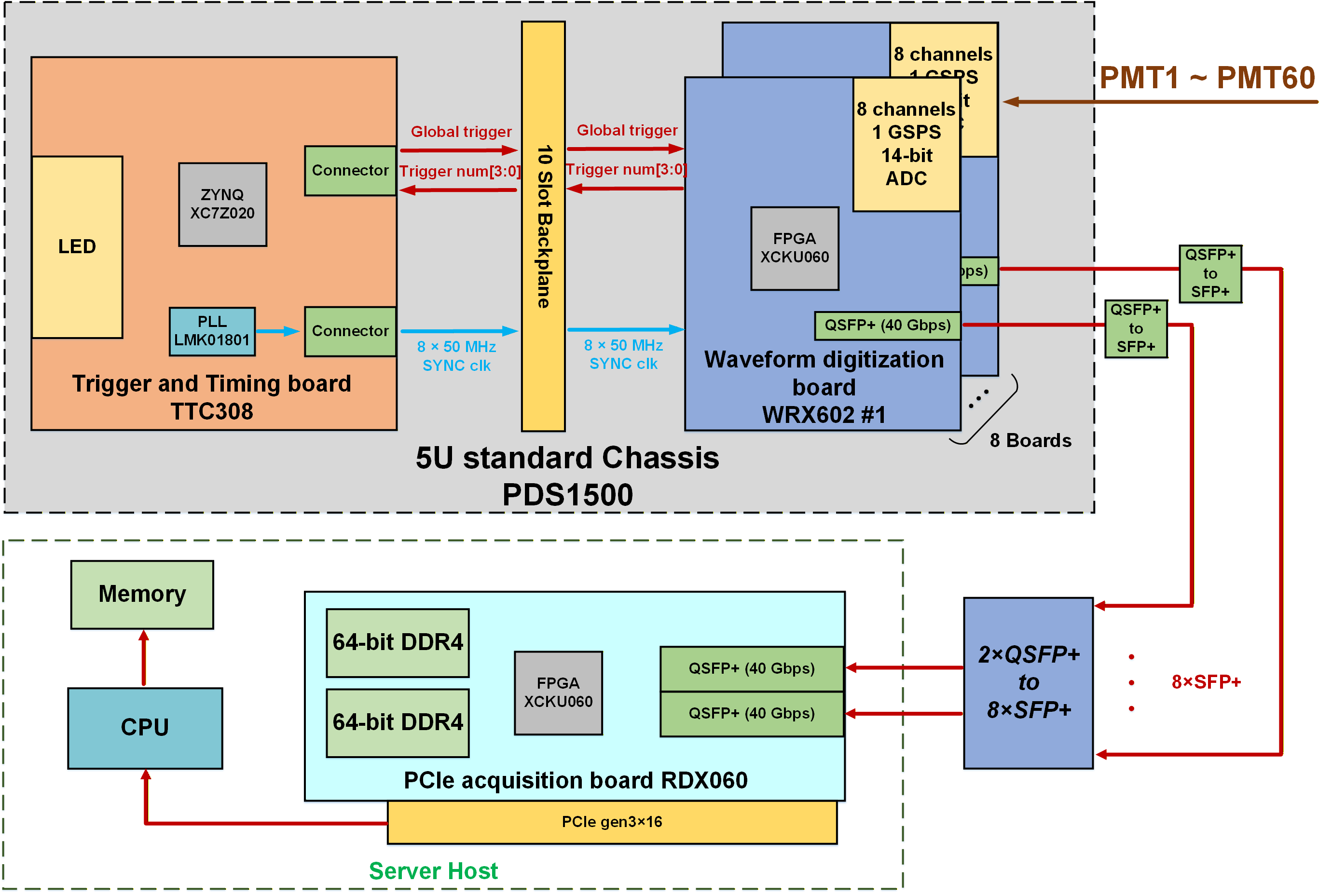}
    \caption{Schematic of the data acquisition system.}
    \label{fig:1tonScheme}
\end{figure}

The schematic diagram of the readout electronics system for the 1 ton prototype of the Jinping Neutrino Experiment is shown in Figure~\ref{fig:1tonScheme}.

\begin{figure}[h][htbp]
    \centering
    \subfigure[ADAM103]{
       \includegraphics[height=0.26\textwidth]{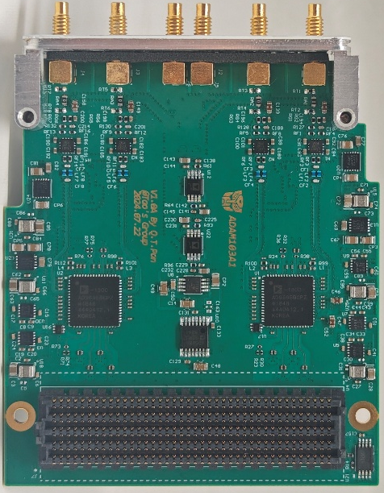}
    }
    \hspace{-3.5mm}
    \subfigure[WRX602]{
        \includegraphics[height=0.26\textwidth]{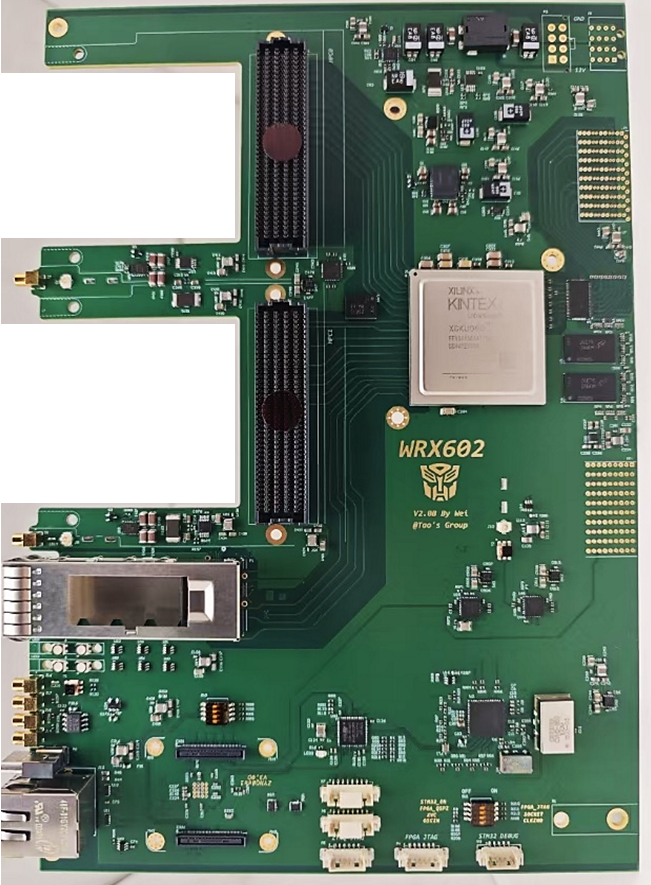}
    }\hspace
    {-3.5mm}
    \subfigure[TTC308]{
        \includegraphics[height=0.26\textwidth]{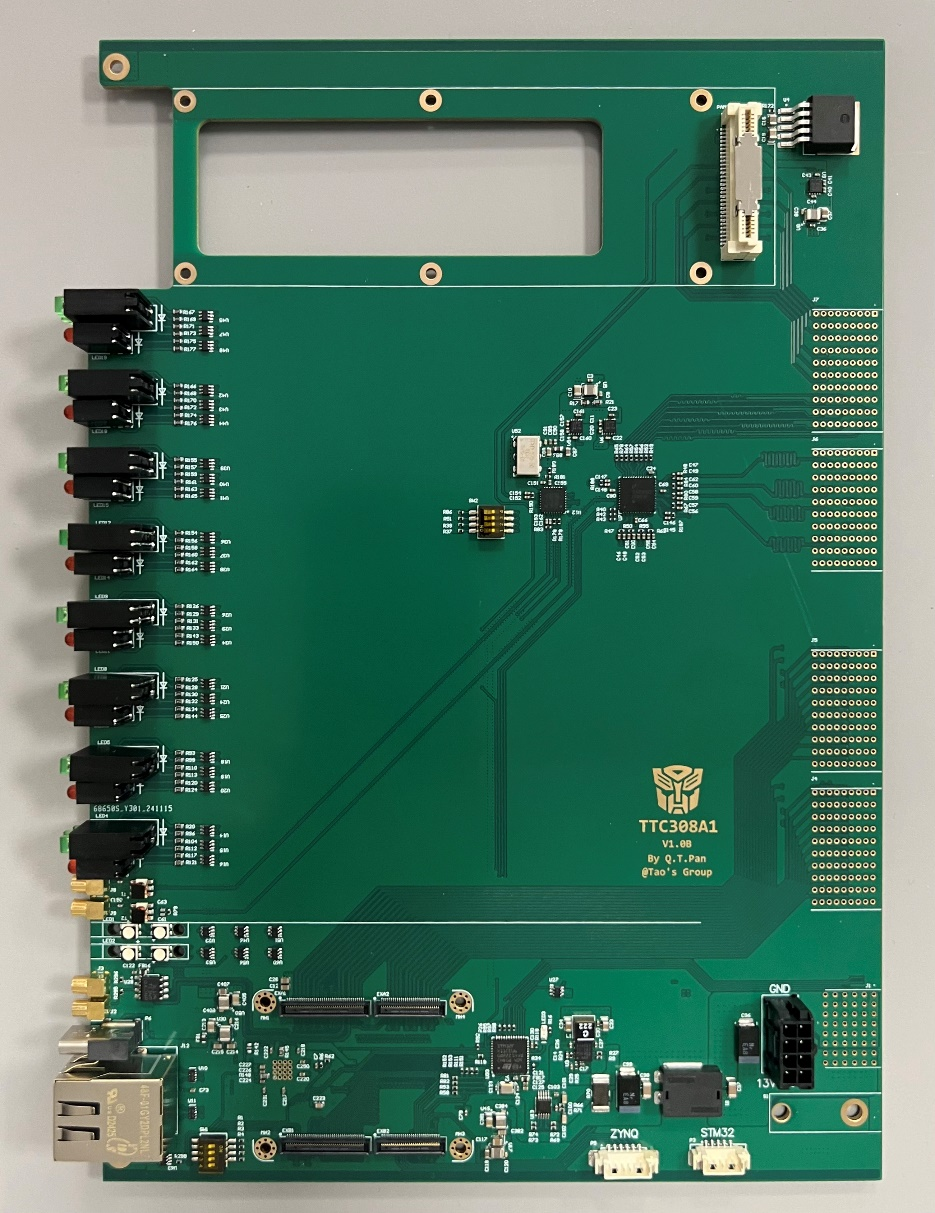}
    }\hspace
    {-3.5mm}
    \subfigure[BP5U10S]{
        \includegraphics[height=0.26\textwidth]{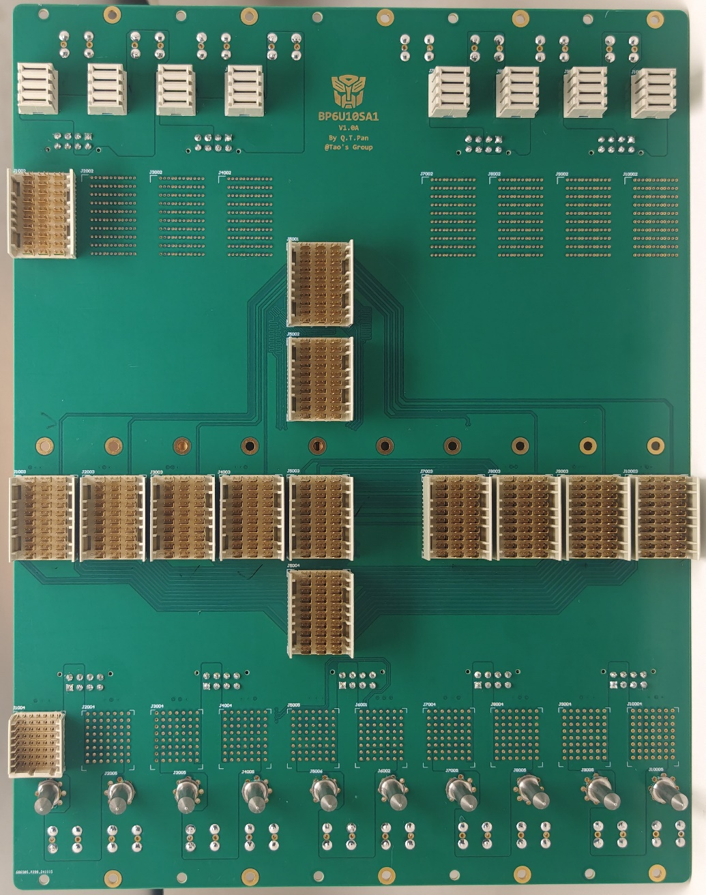}
    }
    \caption{Internal boards. }
    \label{fig:Boards} 
\end{figure}

\begin{figure}[h]
    \centering
    \includegraphics[width=0.5\linewidth]{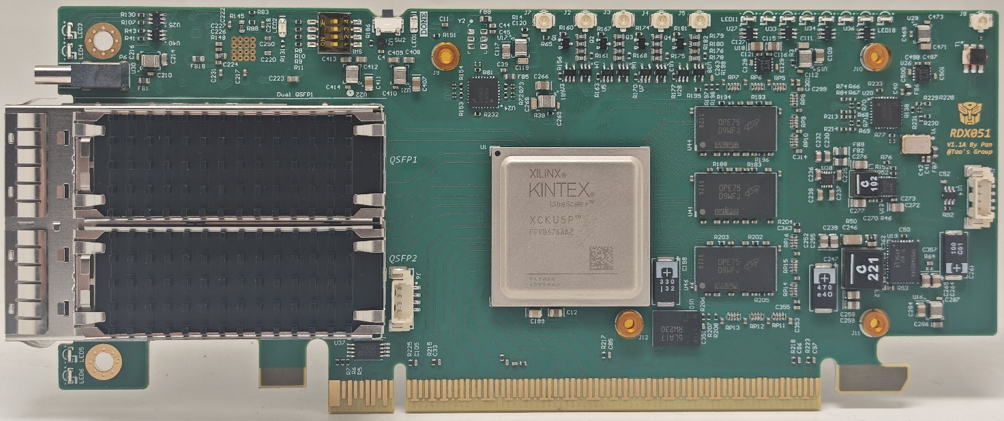}
    \caption{RDX051 PCIe acquisition board.}
    \label{fig:RDX051}
\end{figure}

The output signals from the Photomultiplier Tubes (PMTs) are transmitted directly to the readout electronics via 50 $\Omega$ coaxial cables. The readout boards and the 5U standard cabinet PDS1500 are illustrated in Figure~\ref{fig:Boards}. The system primarily consists of the following components:

\begin{itemize}
    \item \textbf{Eight WRX602 Waveform Digitization Boards:} Each 16-layer PCB is engineered for precise impedance control and high signal integrity. An onboard Xilinx Kintex UltraScale KU060 FPGA~\cite{Xilinx_KU060} manages ADC data buffering, digital pulse shaping, online triggering, and high speed data transmission. Additionally, a Zynq XC7Z020 SoC~\cite{Xilinx_XC7Z020} handles board-level control, status monitoring, and remote firmware updates. Each board integrates two ADAM103 ADC mezzanine cards, providing 8 channels of 14-bit sampling at 1 GSa/s. An onboard LMK04832~\cite{TI_LMK04832} jitter cleaner synchronizes with a 50 MHz backplane clock to generate the 1 GHz ADC sampling clock.
    
    \item \textbf{One TTC308 Trigger and Timing Control Board:} This board aggregates over threshold counts from all eight digitization boards to generate global trigger signals based on programmable logic. It also utilizes an LMK01801~\cite{TI_LMK01801} clock buffer to distribute 50 MHz synchronous clock across the system, ensuring a unified time reference.
    
    \item \textbf{One BP5U10S High Speed Backplane:} Serving as the system backbone, the backplane facilitates high speed data transmission, synchronous clock distribution, and power delivery. It provides one slot for the TTC308 and eight slots for the WRX602 boards. Power is supplied by an HRPG 1000 12~\cite{MeanWell_HRPG1000} switching power supply, distributing 12 V DC to all electronic modules via the backplane.
    
    \item \textbf{One RDX051 PCIe Acquisition Board:} Centered on a Xilinx Kintex UltraScale KU5P FPGA~\cite{Xilinx_KU5P} as shown in Figure~\ref{fig:RDX051}, this board features two front panel QSFP+ interfaces powered by eight GTH transceivers, supporting up to 40 Gbps per interface. Another eight GTH transceivers implement an 8-lane PCIe Gen3 interface, providing a theoretical bandwidth of 64 Gbps. The board also incorporates five high performance storage banks to implement dual 64 bit DDR4 memory channels, offering a total of 64 Gb for data caching.
\end{itemize}

\subsection{Data processing workflow}
\begin{itemize}
\item The FPGA receives 1 GSa/s, 14 bit data streams from eight ADCs on the waveform digitization board. Each data stream is zero-padded to 16 bits to align with a power-of-two data width, simplifying downstream processing. This expanded data is then deserialized into a 250 MHz, 64 bit parallel data stream, where each 64-bit word corresponds to a 4 ns waveform segment.

\item The data from each channel is evaluated against a preset threshold. The number of channels exceeding this threshold is counted and reported to the TTC308 board. The TTC308 board aggregates these over-threshold counts from all eight waveform digitization boards to determine a system-wide total. If this total exceeds a specified limit, a global trigger is simultaneously issued to the eight waveform digitization boards.

\item Upon receiving the global trigger, the data packaging module extracts 1000 ns of waveform data from each ring buffer---specifically, 100 ns of pre-trigger data and 900 ns of post-trigger data. The waveform data from all eight channels, along with the trigger ID and timestamp, are packaged into an event data packet and transmitted to the PCIe acquisition board via an optical fiber link.

\item The PCIe acquisition board receives the data packets from the eight waveform digitization boards via optical fiber and buffers them in its onboard DDR4 memory. Two 4-Gb memory blocks are allocated on the board to operate as ping-pong buffers. Data transfer and control between the PCIe acquisition board and the host PC are handled by an 8-lane PCIe Gen3 link. The host monitors the PCIe board's internal buffers; when one of the 4-Gb buffers fills up, the host issues a transfer command. The data is then transmitted over the PCIe link, and the host saves it to the hard disk in ROOT format.
\end{itemize}

\begin{figure}[h]
\centering
\includegraphics[width=.9\textwidth]{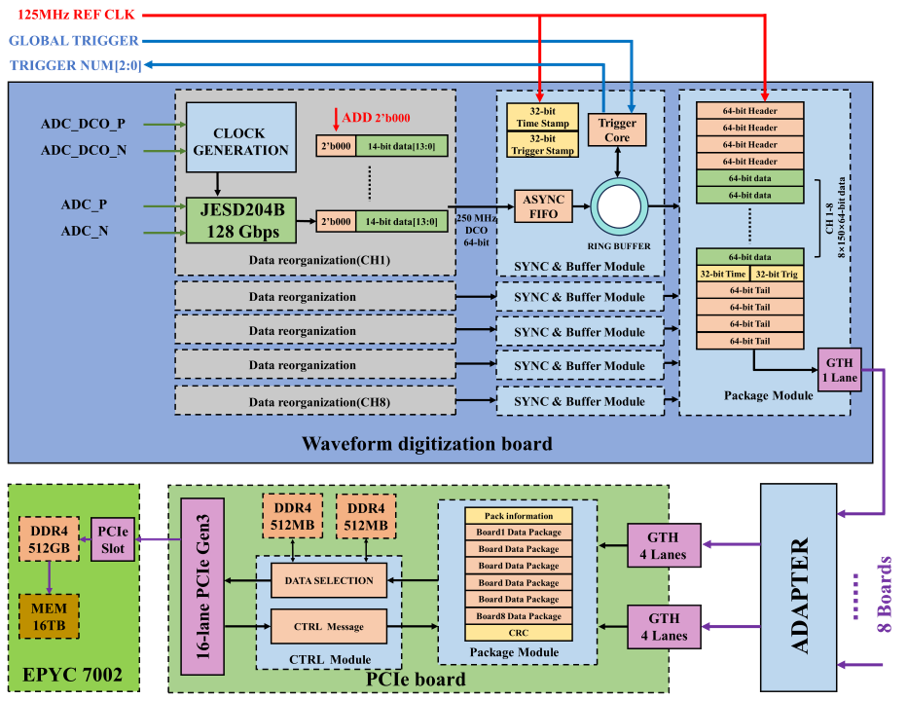}
\caption{The workflow of the 60 channels electronics system.\label{data workflow}}
\end{figure}

\subsection{Clock distribution architecture}
\begin{figure}[h]
    \centering
    \includegraphics[width=0.49\linewidth]{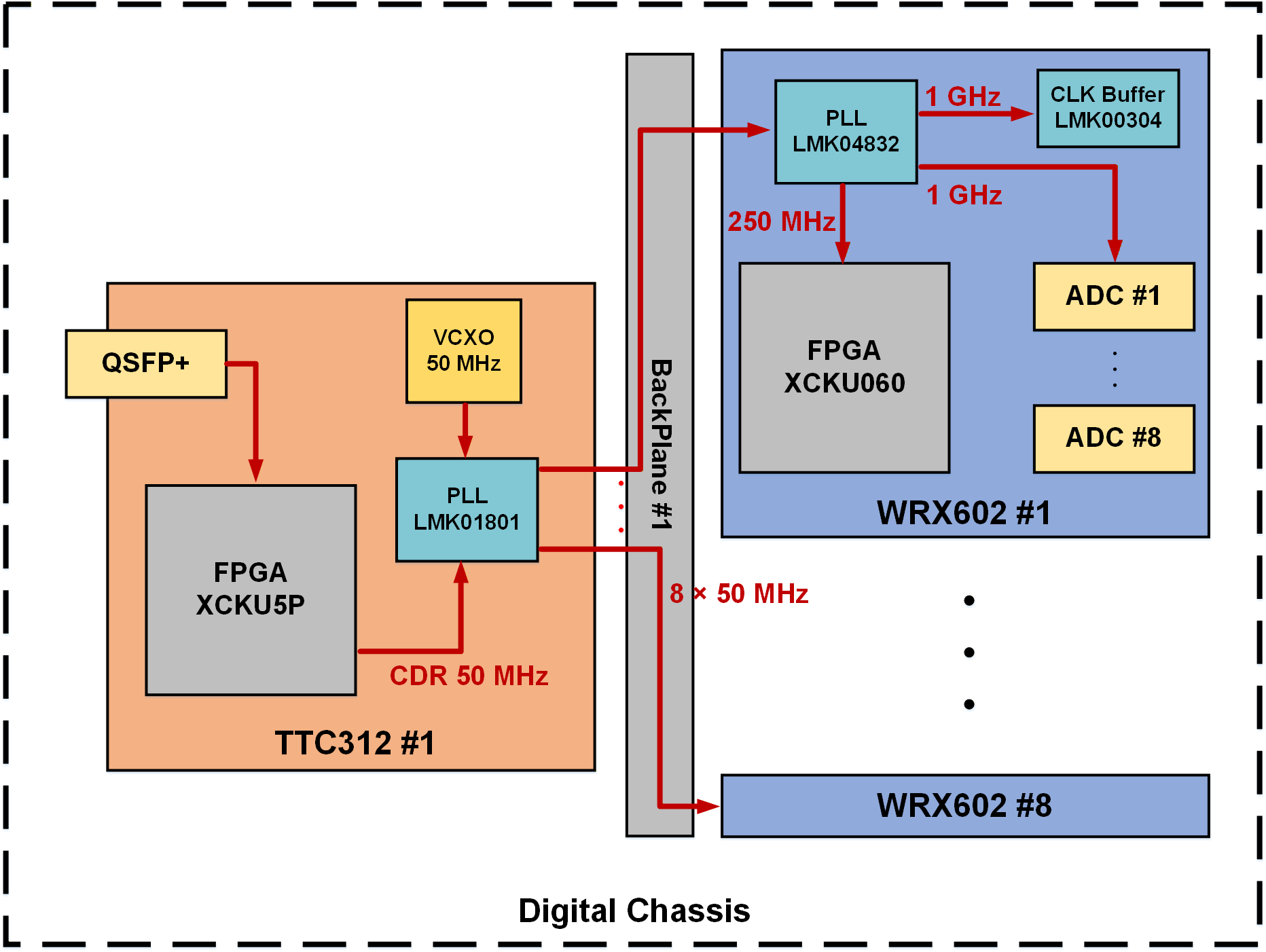}
    \caption{The clock distribution architecture of the 60-channel electronics system.}
    \label{fig:1tonclk}
\end{figure}

Time resolution fundamentally dictates the precision of vertex reconstruction in the Jinping Neutrino Experiment; consequently, the readout electronics must achieve and maintain stringent inter-channel synchronization. The proposed clock distribution architecture designed to meet this requirement is illustrated in Figure~\ref{fig:1tonclk}. 

At the system level, the TTC308 timing board utilizes an onboard Voltage-Controlled Crystal Oscillator (VCXO) and an LMK01801 clock buffer to distribute a 50~MHz synchronous reference to all eight digitization boards within a single crate. On each individual digitizer board, an LMK04832 jitter cleaner multiplies this reference to generate both a 250~MHz user clock for FPGA logic and a 1~GHz sampling clock for the ADCs. By operating the LMK04832 in Zero-Delay Mode (ZDM)—where the 250~MHz output is deterministically fed back to the second-stage Phase-Locked Loop (PLL)—the phases of all derived clocks remain strictly aligned with the incoming 50~MHz reference. This architecture minimizes phase skew and guarantees deterministic latency across the entire crate. 

This deterministic clock architecture fundamentally resolves the inter-board synchronization ambiguities commonly encountered in commercial legacy systems. For instance, multi-board synchronization in systems like the CAEN V1751 typically relies on daisy-chained front-panel trigger distributions or reference clocks without strict phase-feedback loops. Such topologies inherently suffer from phase ambiguity—often introducing a $\pm 1$ sampling clock cycle ($\pm 1$~ns at 1~GSa/s) synchronization uncertainty—and are susceptible to thermal drifts, requiring post-processing software timestamp alignment. 

In contrast, the proposed system achieves absolute trigger consistency across all distributed channels. Coupled with the zero-delay clock network, the global trigger signal is distributed via the backplane using an equal-length star topology. This ensures that the trigger assertion aligns deterministically with the exact same phase of the 1~GHz sampling clock on every digitization board. Consequently, the system guarantees true simultaneous waveform acquisition across all 60 channels without the need for post-acquisition software "cropping" or alignment, significantly reducing offline data processing overhead and eliminating inter-board timing jitter.

\section{Performance test}
\subsection{The JNE-1ton prototype detector}
The JNE 1-ton prototype was constructed to validate key technologies for the JNE, including the slow liquid scintillator~\cite{Guo:2017nnr}, the 8-inch MCP-PMT photomultiplier tubes\cite{Zhang:2023ued}, and especially the new electronic systems. This prototype detector is located in the CJPL-I, with a rock coverage depth of 2400 meters, providing an extremely low cosmic ray background environment. From 2017 to 2023, the prototype successfully completed the first stage of data collection tasks, measuring the background concentrations of natural radioactive nuclides such as $^{238}$U, $^{232}$Th, and $^{40}$K in the laboratory environment~\cite{wu2023performance}, as well as the precise value of the underground muon flux: $(3.56\pm0.16_{\text{stat}}\pm0.10_{\text{syst}}) \times 10^{-10}\,{\text{cm}^{-2}\cdot s^{-1}}$~\cite{PhysRevD.110.112017}.

\begin{figure}[h]
    \centering
    \includegraphics[width=0.85\linewidth]{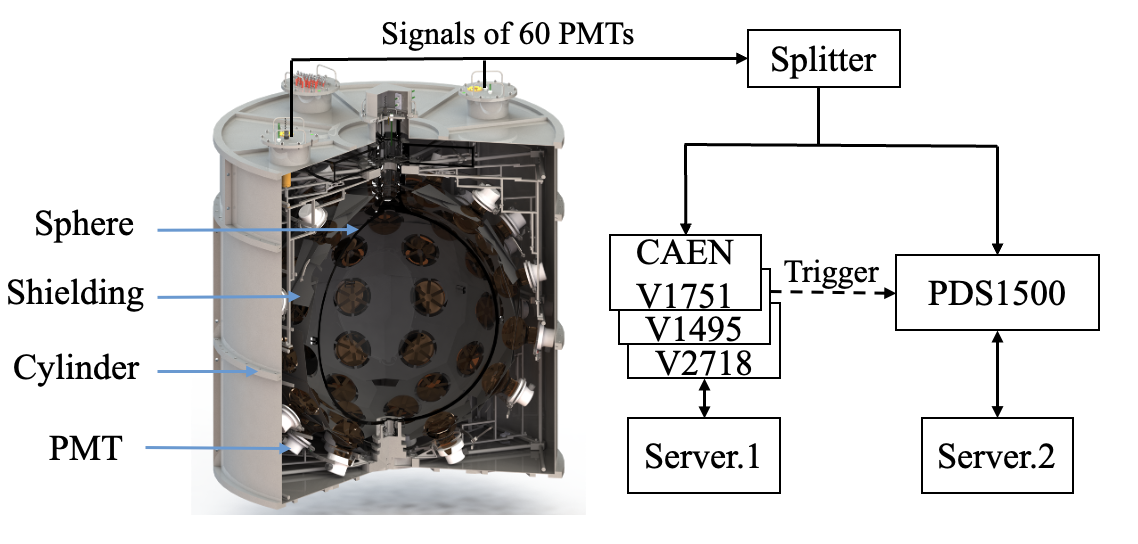}
    \caption{The structure of upgraded one-ton prototype detector and the connection method with the CAEN V1751 system and PDS1500 system (the new self-developed electronics).}
    \label{fig:upgraded}
\end{figure}

Beginning from 2023, we carried out a critical upgrade for the prototype detector: all the PMTs were replaced with 8-inch MCP-PMTs developed by Northern Night Vision Technology Co., Ltd.~\cite{Zhang:2023ued}, increasing the total number from 30 to 60. Simultaneously, the PMT supporting and shielding structures were reconfigured (Figure~\ref{fig:upgraded})~\cite{wu2023performance,Yang:2024yco}. After the upgrade, the central medium of the detector supports two operation modes: pure water and slow liquid scintillator~\cite{Guo:2017nnr}. 
During the test of electronic system data acquisition, the data were collected in the slow liquid scintillator phase, which was designed to evaluate the performance of the new electronic system in a real experimental environment. Unfortunately, 7 out of 60 PMTs (channels) were damaged, leaving only 53 channels that could actually be used for data collection.

The upgraded prototype still used the CAEN V1751 system for continuous data acquisition during the early stage of electronic testing: the FADC board V1751 was increased from 4 cards (32 channels) to 8 cards (64 channels); the trigger module (V1495), data transmission link (V2718 + A2818 PCIe card), and server writing program remained unchanged~\cite{wu2023performance}.


To avoid affecting the existing data acquisition scheme, we adopted the connection method shown in Figure~\ref{fig:upgraded}: the signal from each PMT is split into two paths by a splitter, which are then fed into the CAEN system and the PDS1500 system, respectively. The splitter ensures that the waveforms of the two output signals are consistent with that of the single input signal. The two systems are controlled independently by separate servers and do not interfere with each other. Meanwhile, the PDS1500 has the optional capability to receive the trigger signal from the CAEN system.

\subsection{Systems specification comparison}
To contextualize the performance of the proposed design, Table~\ref{tab:digitizer_comparison}  presents a comprehensive comparison of key specifications between the PDS1500 and the commercially established CAEN V1751 digitizer. Critical parameters— including sampling rate, resolution, effective number of bits (ENOB), and analog bandwidth — are detailed to highlight the relative advantages and operational trade-offs of each system. The PDS1500 digitizer offers an analog bandwidth of 350 MHz. Given the detector’s intrinsic rise time of ~5 ns, the signal spectrum is well within this bandwidth (required $BW \approx 0.35 / t_r \approx 70 \text{ MHz}$). Consequently, the 350 MHz cutoff ensures no measurable signal distortion or timing degradation compared to higher-bandwidth alternatives.

\begin{table}
    \centering
    \caption{Comprehensive Comparison of Key Specifications between CAEN V1751 and PDS11500. The values of ENOB (Effective number of bits) and the maximum trigger frequency (Max. Trigger Frequency) are both derived from the actual measurement results presented in the following sections.}
    \label{tab:digitizer_comparison}
    \begin{tabular}{lcc}
        \toprule
        \textbf{Parameter} & \makecell{\textbf{CAEN V1751} \\ (Previous)}  & \makecell{ \textbf{PDS1500 WRX602}\\ (New)} \\
        \midrule
        \multicolumn{3}{l}{\textit{Architecture \& Physical Specifications}} \\
        Form Factor & VME64 & CPCI \\
        Number of Channels & 8 (4 in DES mode) & 8 \\
    
        \midrule
        \multicolumn{3}{l}{\textit{Analog Front-End Characteristics}} \\
        Sampling Rate & 1 GSa/s & 1 GSa/s \\
        Analog Bandwidth ($-3$ dB) & 500 MHz & 350 MHz \\
        Input Voltage Range & 1.0 Vpp  & 0.8 Vpp(adjustable) \\
        Input Impedance & 50 $\Omega$ & 50 $\Omega$ \\
        \midrule
        \multicolumn{3}{l}{\textit{Dynamic Performance Metrics}} \\
        Physical Resolution & 10-bit & 14-bit \\
        ENOB & 9.0 bit & 10.6 bit \\
        SNR & $\sim$56.2 dB & $\sim$65.6 dB \\
        \midrule
        \multicolumn{3}{l}{\textit{Triggering \& Data Throughput}} \\
        Max. Trigger Frequency & 0.46 kHz & 50 kHz \\
        On-board Memory Depth & 1.83 MS/ch  & 500 MS/ch \\
        Multi-board Sync & Supported (Daisy chain topology) & Supported (Star topology) \\
        \bottomrule
    \end{tabular}
\end{table}

\subsection{Performance evaluation}
To evaluate the performance of the PDS1500 system, this paper conducts a comparative analysis with the previous system in terms of six key aspects: data loss rate, baseline stability, waveform fidelity, timing drift, detector energy threshold, and energy resolution, thereby verifying whether the system meets the design requirements of the JNE detector.

To facilitate the measurement and comparison of the above performance metrics, corresponding data acquisition strategies are adopted for different indicators. When comparing the data loss probability, baseline stability, and waveform quality of the two systems, a synchronous data acquisition mode was used, in which the PDS1500 receives trigger signals from the CAEN V1751 system. 
When measuring the time drift between different channels within the same system, an LED calibration data acquisition mode was employed. 
When comparing detector thresholds, a natural radioactivity data acquisition mode was adopted. 
When comparing energy resolution, an AmBe calibration source was placed at the center of the detector in the corresponding data acquisition mode.

\subsubsection{Data loss}
We obtained two datasets of two electronic systems in the external trigger mode of CAEN system (see Table~\ref{tab:TwoSystemDaq}). 
Statistical analysis of over a million events shows that the PDS1500 system achieved zero event loss and captured two additional events that were not recorded by CAEN. 
\begin{table}
    \centering
    \caption{The two sets of synchronous data are respectively from the CAEN V1751 and the PDS1500 systems, the latter of which is provided with an external trigger signal by the CAEN V1495 trigger board.}
    \begin{tabular}{ccccc}
        \toprule
         Term & CAEN V1751 system   & PDS1500 system  \\
        \midrule
         DAQ.1& 409598 & 409600 &  \\
         DAQ.2& 614400 & 614400 &  \\
        \bottomrule
    \end{tabular}
    \label{tab:TwoSystemDaq}
\end{table}

The acquisition time window of the V1751 is approximately 1029 ns in the 1-ton detector system, while the recorded time window is 980 ns for the 8 cards trigger synchronization. There is a dead time of 49 ns. 
In contrast, the PDS1500 has an acquisition time window of 1000 ns and does not require multi‑board waveform cropping; therefore, it is able to record signals in the situations described above.

Thanks to its stable internal clock distribution architecture, the PDS1500 has almost zero dead time during acquisition over a $\sim$1 $\mu$s window, and experimental measurements indicate that its data loss performance is comparable to (i.e., no worse than) that of the CAEN V1751 system.

\subsubsection{Baseline stability}
\begin{figure}[h]
    \centering
    \subfigure[CAEN V1751]{
    \includegraphics[width=0.45\linewidth]{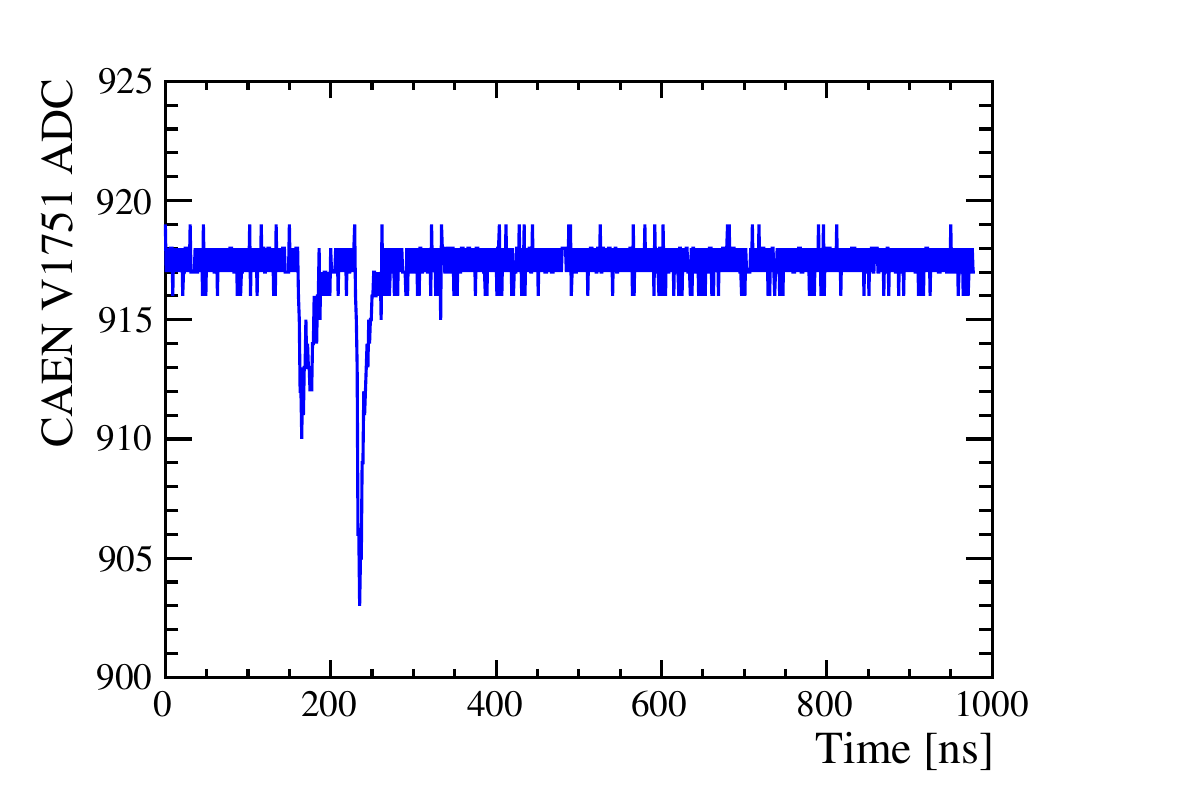}
    }
    \subfigure[PDS1500]{
    \includegraphics[width=0.45\linewidth]{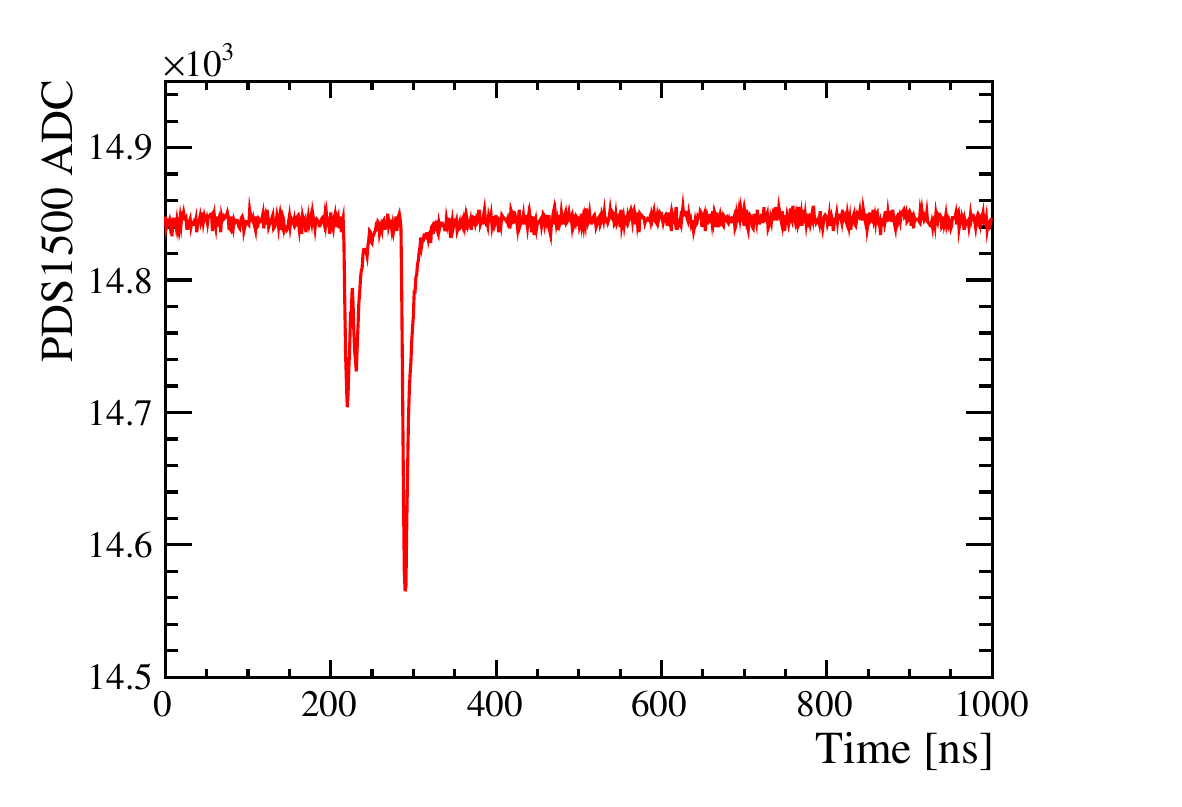}
    }
    \caption{The waveforms of the same signal acquired on the CAEN V1751 and PDS1500 systems, the discrepancy in Y-axis is caused by the difference in their physical resolutions.}
    \label{fig:time-window}
\end{figure}

Due to the different physical resolution in the two electronic data acquisition systems, the original amplitude count values obtained for the same trigger event and the same channel are not directly comparable, as shown in Figure~\ref{fig:time-window}. 

\begin{figure}[h]
    \centering
    \subfigure[Single channel]{
    \includegraphics[width=0.45\linewidth]{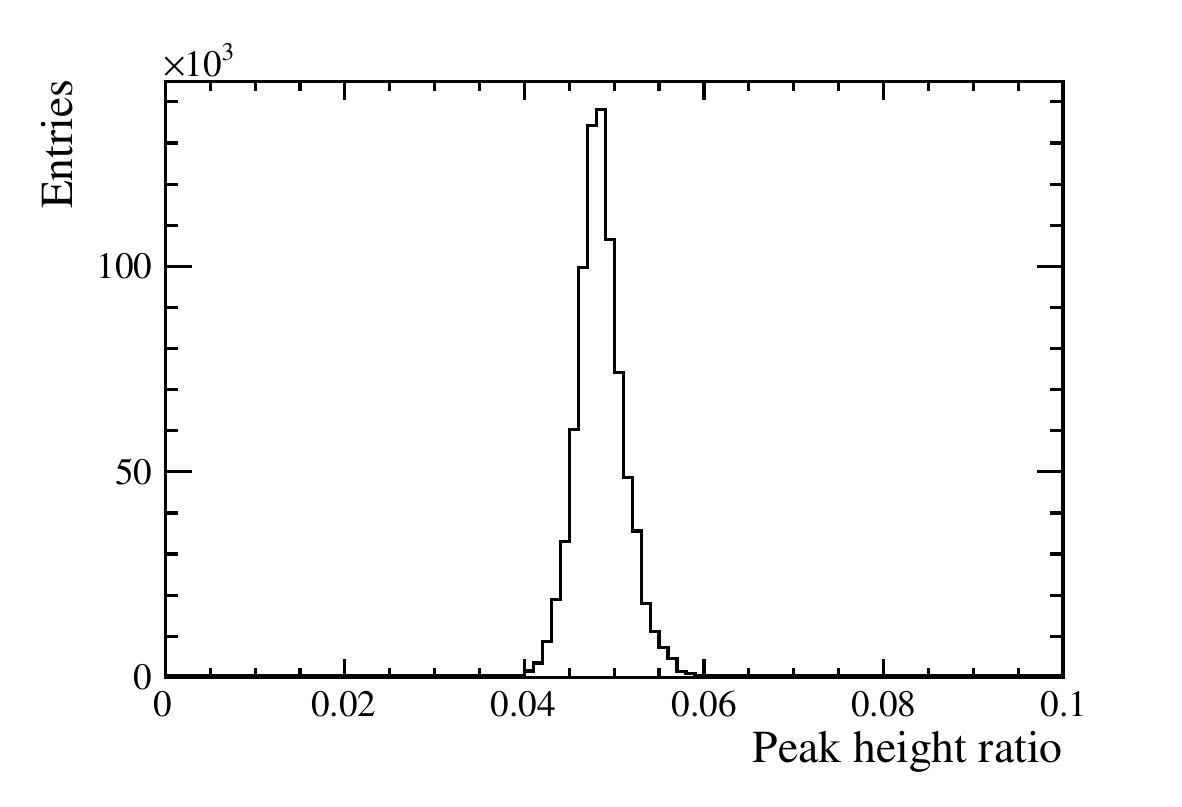}
    }
    \subfigure[53 channels]{
    \includegraphics[width=0.45\linewidth]{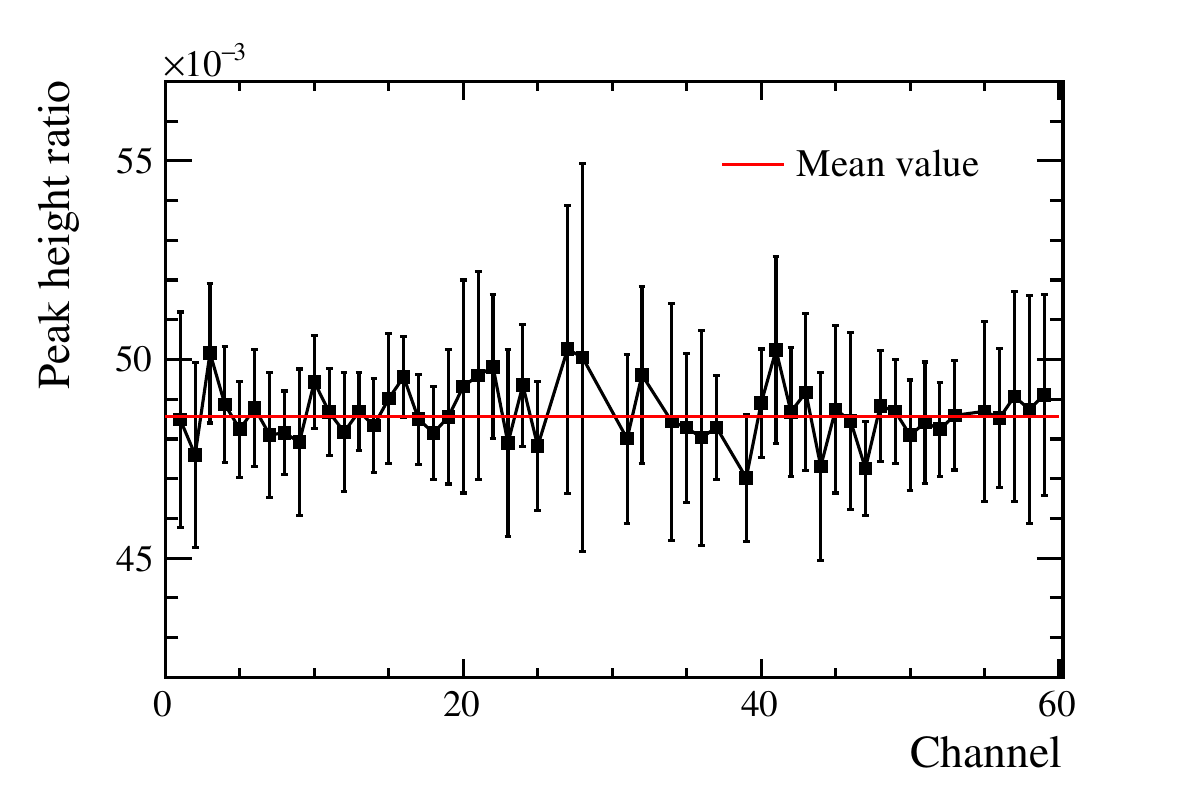}
    }
    \caption{The distribution of peak height ratio values for each event in the CAEN V1751 and PDS1500 systems for a single channel, as well as the mean values of ratio and their standard deviations corresponding to all 53 channels.}
    \label{fig:caenScale}
\end{figure}

To quantitatively compare the data in different systems, based on the statistical analysis of the distribution of peak height ratios (where peak height is defined as the difference between the signal minimum value and the baseline) of the two systems, the results are shown in Figure~\ref{fig:caenScale}. Thus, the normalization factor of the PDS1500 system relative to the CAEN V1751 system is calibrated to 0.04858.

\begin{figure}[h]
    \centering
    \includegraphics[width=0.5\linewidth]{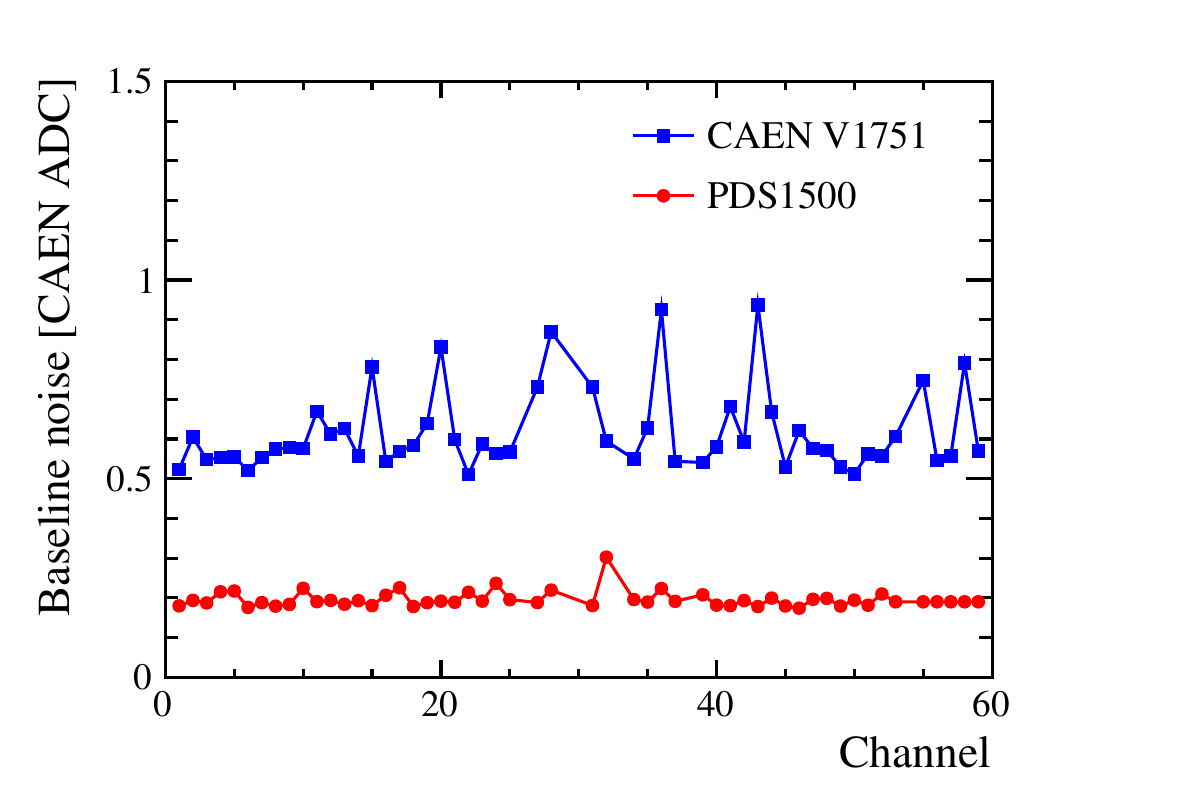}
    \caption{Baseline noise of CAEN V1751 and PDS1500 systems of different channels. The noise level of the PDS1500 system is only one third that of the CAEN V1751 system.}
    \label{fig:baselinestd}
\end{figure}

Baseline stability, measured by single-channel baseline noise (defined as the standard deviation of ADC values in a 50 ns window), directly affects measurement accuracy.
As shown in Figure~\ref{fig:baselinestd}, the baseline noise level of the PDS1500 system is approximately one-third that of the CAEN V1751 system after normalization, indicating a significant improvement in baseline stability.

A stable baseline reduces measurement bias caused by random fluctuations and improves the accuracy of waveform measurement. This is especially critical when using single‑channel fixed ADC threshold triggering. A stable baseline also helps avoid trigger instability or missed triggers that may occur when the waveform maximum lies exactly at the edge of the ADC threshold.

\subsubsection{Waveform quality}
When performing waveform analysis and comparison, waveform alignment is essential. Figure~\ref{fig:time-window} shows inconsistencies in baseline, amplitude and time between their original waveforms.

\begin{figure}[h]
    \centering
    \subfigure[Waveform alignment]{
    \includegraphics[width=0.45\linewidth]{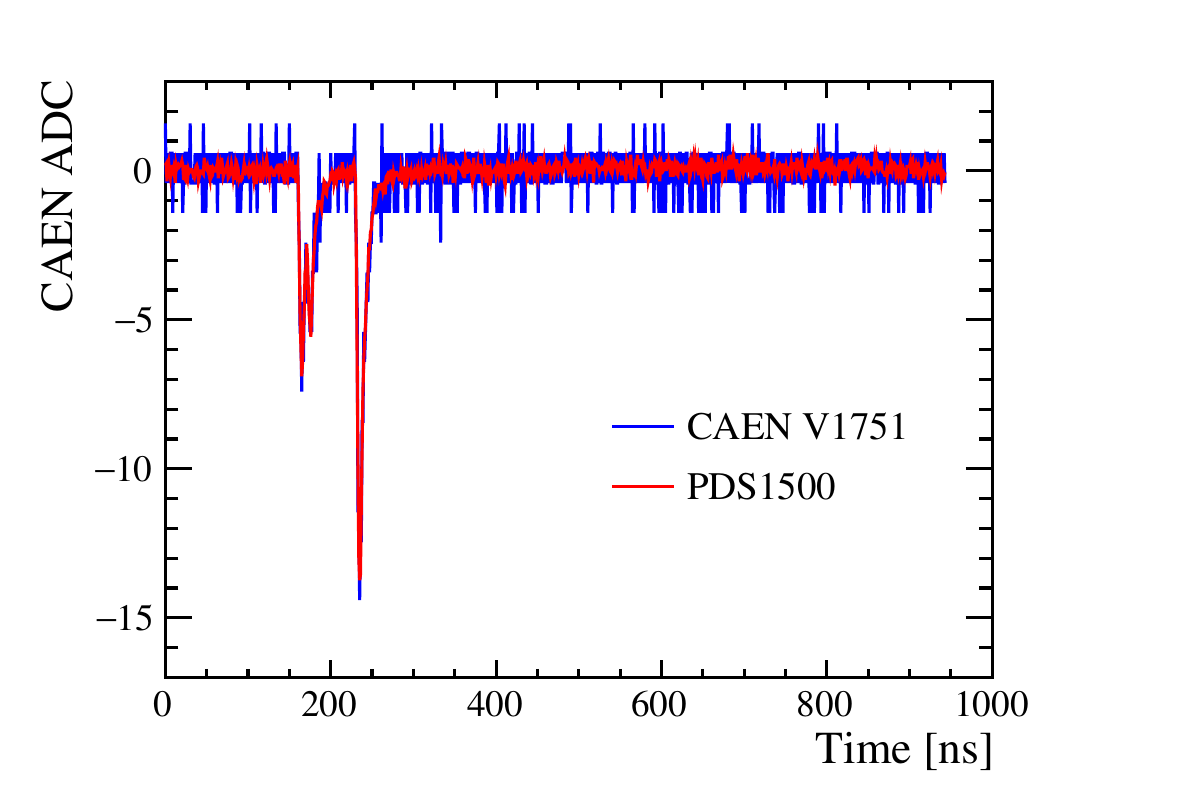}
    }
    \subfigure[Differences]{
    \includegraphics[width=0.45\linewidth]{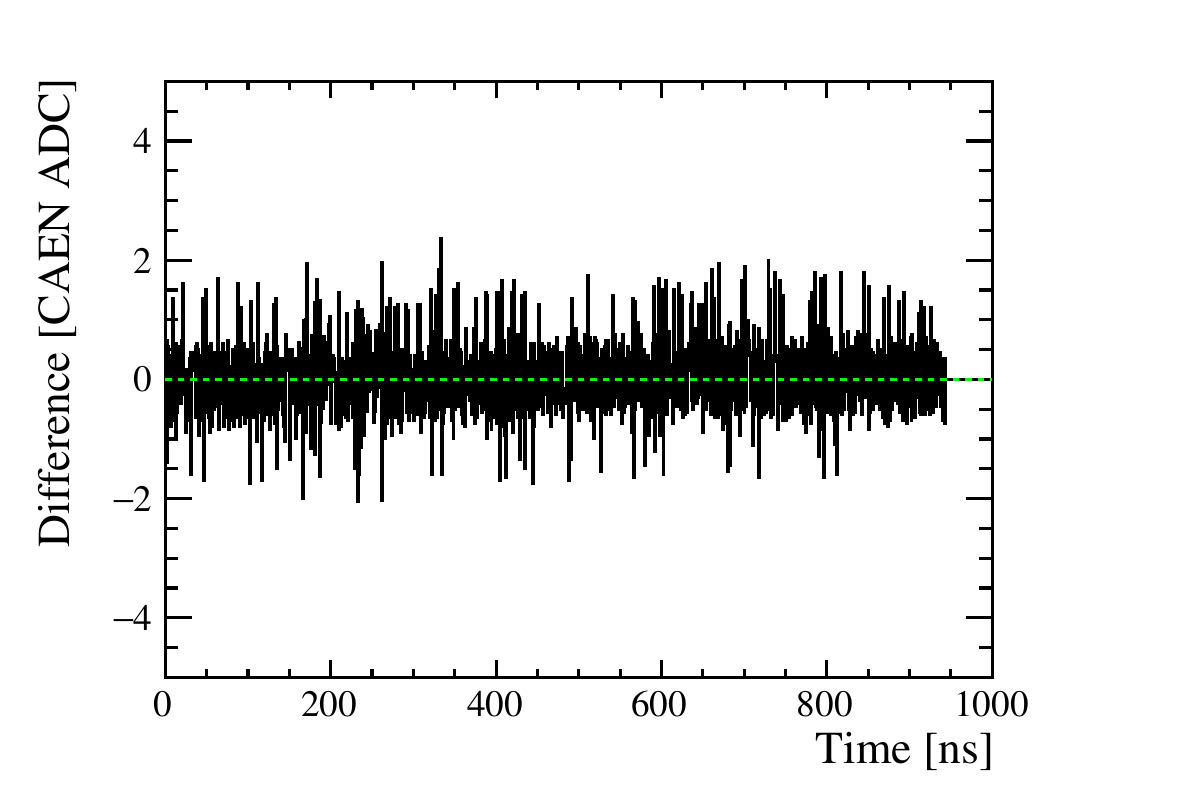}
    }
    \caption{ Example of PMT waveform alignment between CAEN V1751 and PDS1500 systems, as well as their differences.}
    \label{fig:wavealigned}
\end{figure}

The baseline alignment is completed by subtracting the baseline from the waveforms.
The amplitude alignment used the previously determined value of 0.04858 to scale the waveform of the PDS1500.
The time misalignment is due to different trigger time and different transition time in the two electronics systems.
Time alignment employs an edge-synchronous strategy: 
first, identify the main peak, which is the global minimum point in the waveform; 
then, position the rising edge as the first sampling point that is 20\% lower than the height of this peak; 
subsequently, align the rising edge points of the corresponding waveforms from the two systems strictly to the same time coordinate. 
Figure~\ref{fig:wavealigned} shows an example of aligned waveforms and their difference, which correspond to the waveforms shown in Figure~\ref{fig:time-window}.

\begin{figure}[h]
    \centering
    \subfigure[Single channel]{
    \includegraphics[width=0.45\linewidth]{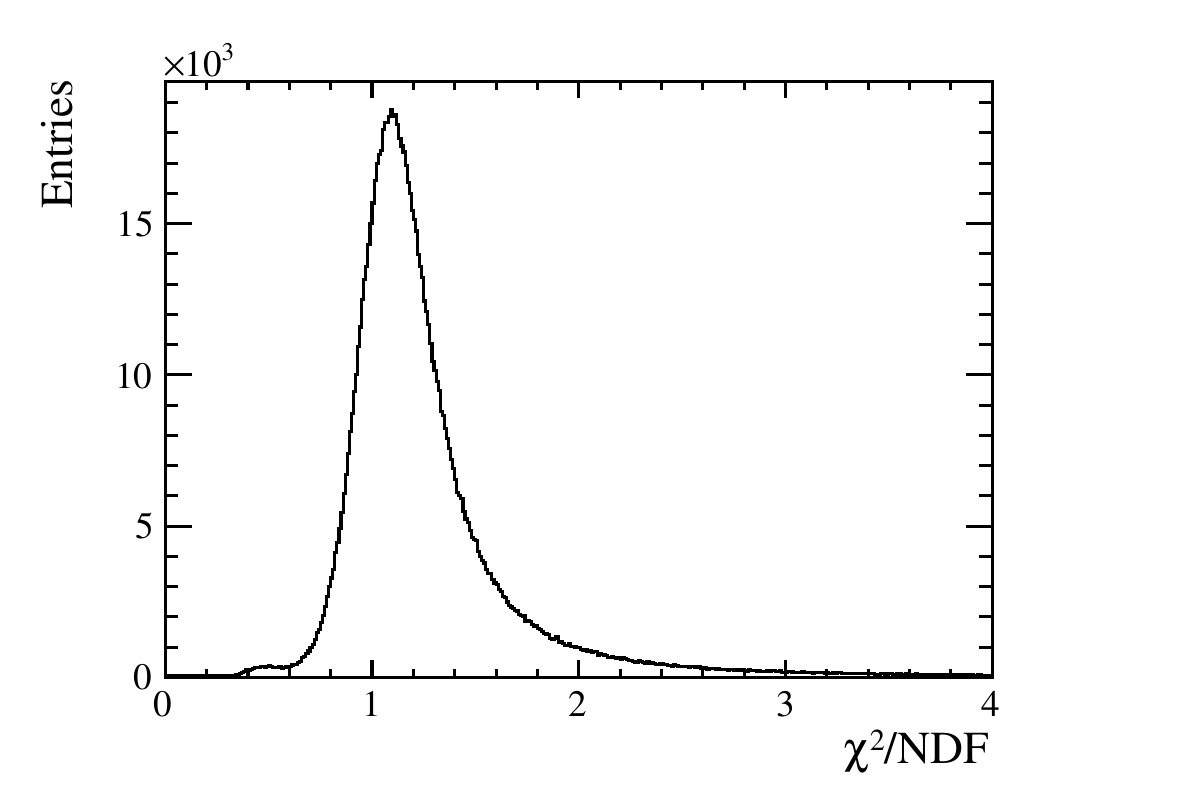}
    }
    \subfigure[53 channels]{
    \includegraphics[width=0.45\linewidth]{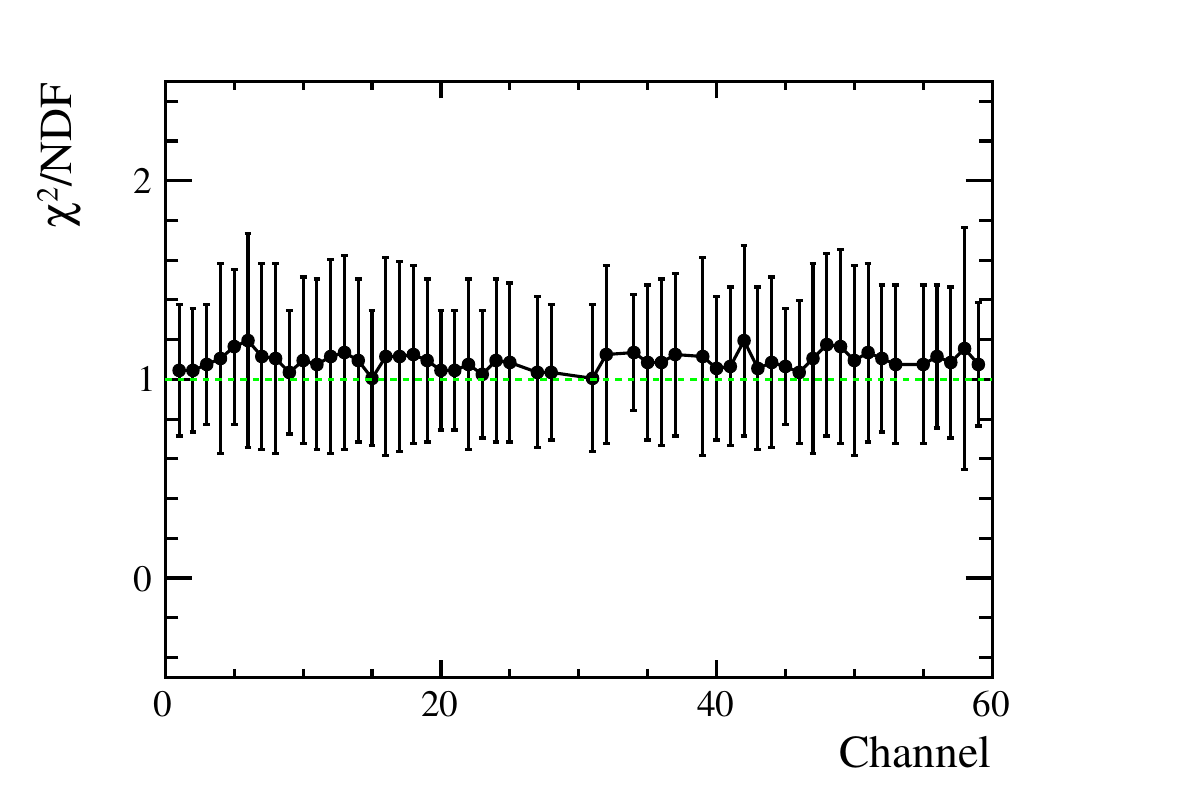}
    }
    \caption{The distribution of $\chi^{2}$/NDF for the 10-th channel, as well as the peak positions of $\chi^{2}$/NDF distributions for different channels, with the error expressed in terms of the peak half-width.}
    \label{fig:chi2diss}
\end{figure}

To quantitatively evaluate the consistency of the waveform acquisition of the two electronic systems, we use over a million identical physical events that are simultaneously triggered in the two systems to conduct point-by-point comparison of their original waveforms; on this basis, the $\chi^2$ defined as follows:
\begin{equation}
	\chi^2 = \sum_t \frac{[w_1(t) - w_2(t)]^2}{\sigma_1^2 + \sigma_2^2},
    \label{eq:Chi2ForWave}
\end{equation}
where $w(t)$ refer to wave values in time $t$ from CAEN V1751 or PDS1500 systems after the wave alignments, and $\sigma$ refer to their baseline noise before.
With the $\chi^2$ calculation, waveform discrepancies between the two electronic systems are quantified across all events. 
Figure~\ref{fig:chi2diss} shows the $\chi^2$/NDF distribution for Channel 10, and displays the peak positions of $\chi^2$/NDF distribution across the 53 live channels. 
All the peak positions near 1.0, indicating consistency with expected statistical fluctuations and thus confirming good agreement in the waveform.  

By comparing with the waveforms from the CAEN V1751 system, which has a lower physical resolution (10 bits), theoretically, the PDS1500 has a higher value (14 bit), and the baseline noise is only one third of the CAEN V1751 system, so the PDS1500 waveform is expected to more faithfully reproduce the true signal, revealing finer pulse details. 

\subsubsection{Time drift magnitude}
In the detector, the PMT, signal cables, and electronic circuits will introduce inherent time offsets among channels; if this offset undergoes uncontrollable drift over time, it will significantly deteriorate the time resolution performance and the accuracy of event reconstruction.

\begin{figure}[h]
     \centering
     \subfigure[LED calibration]{
     \includegraphics[width=0.35\linewidth]{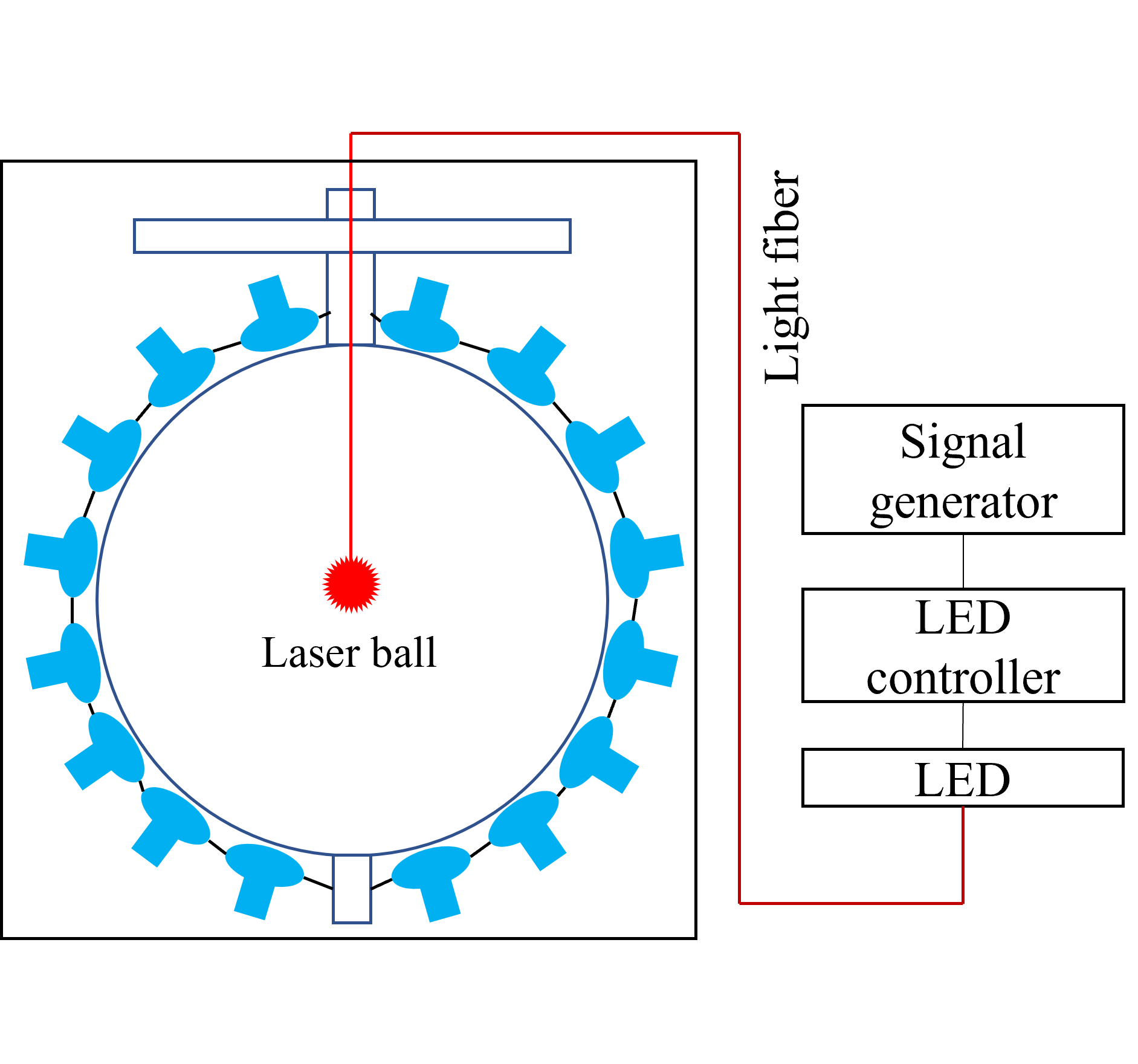}
     }
     \subfigure[Waveform from the calibration data]{
      \includegraphics[width=0.60\linewidth]{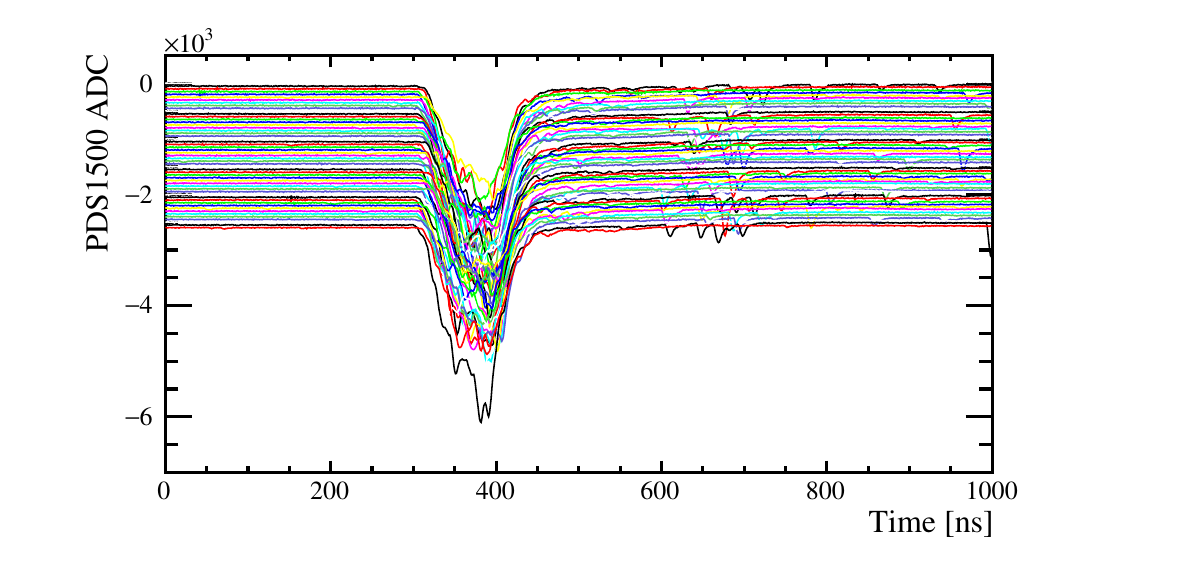}
      }
     \caption{Schematic diagram of the detector time calibration (left), and the signal waveforms from the 53 live PMTs collected by the PDS1500 electronics system (right). The baseline has been optimized for demonstration purposes.}
\label{fig:DiffusionBall}
\end{figure}

During the time calibration process, the 415 nm LED light source driven by the signal generator is coupled through an optical fiber to a polytetrafluoroethylene laser ball~\cite{Ouyang:2025phk} located at the center of the detector (Figure~\ref{fig:DiffusionBall}). 
The LED emission pulse width is approximately 100 ns, and its light intensity is sufficient to ensure that all photomultiplier tubes can receive measurable synchronous light signals.

\begin{figure}[h]
    \centering
    \subfigure[Charge integration distribution ]{
    \includegraphics[width=0.45\linewidth]{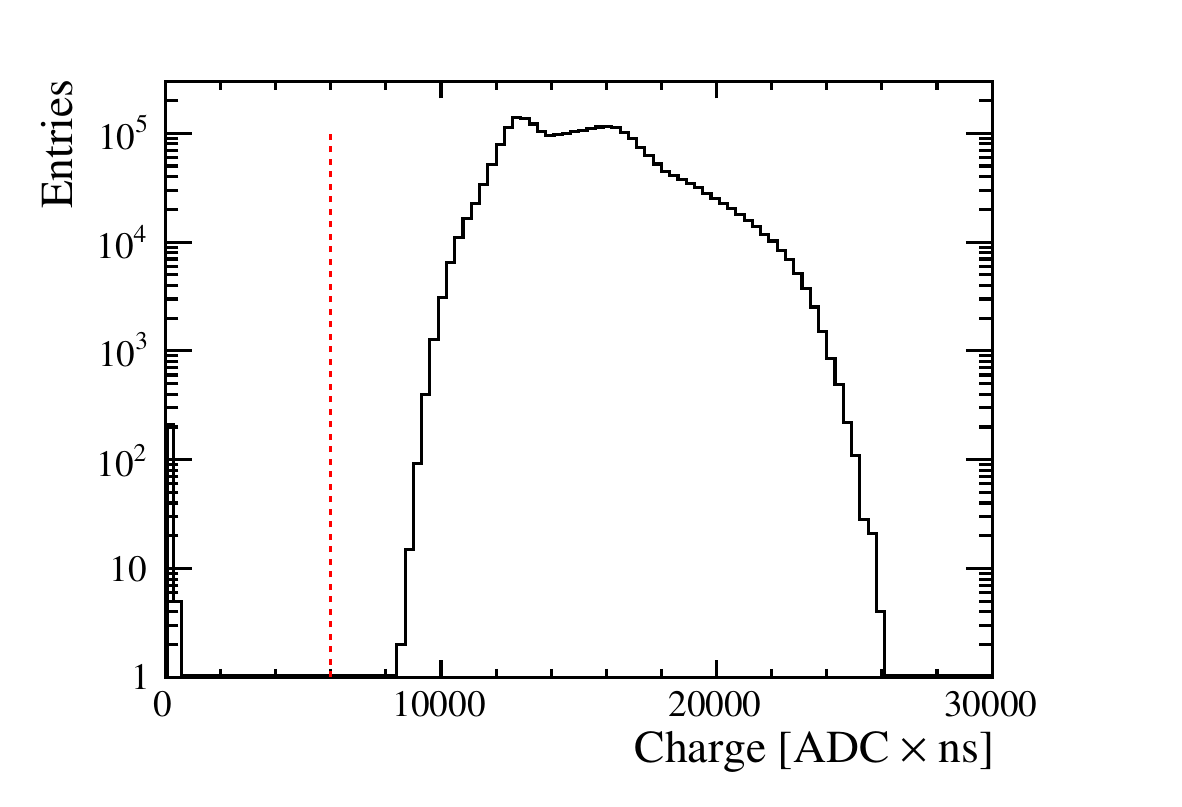}
    }
    \subfigure[Rise time difference]{
        \includegraphics[width=0.45\linewidth]{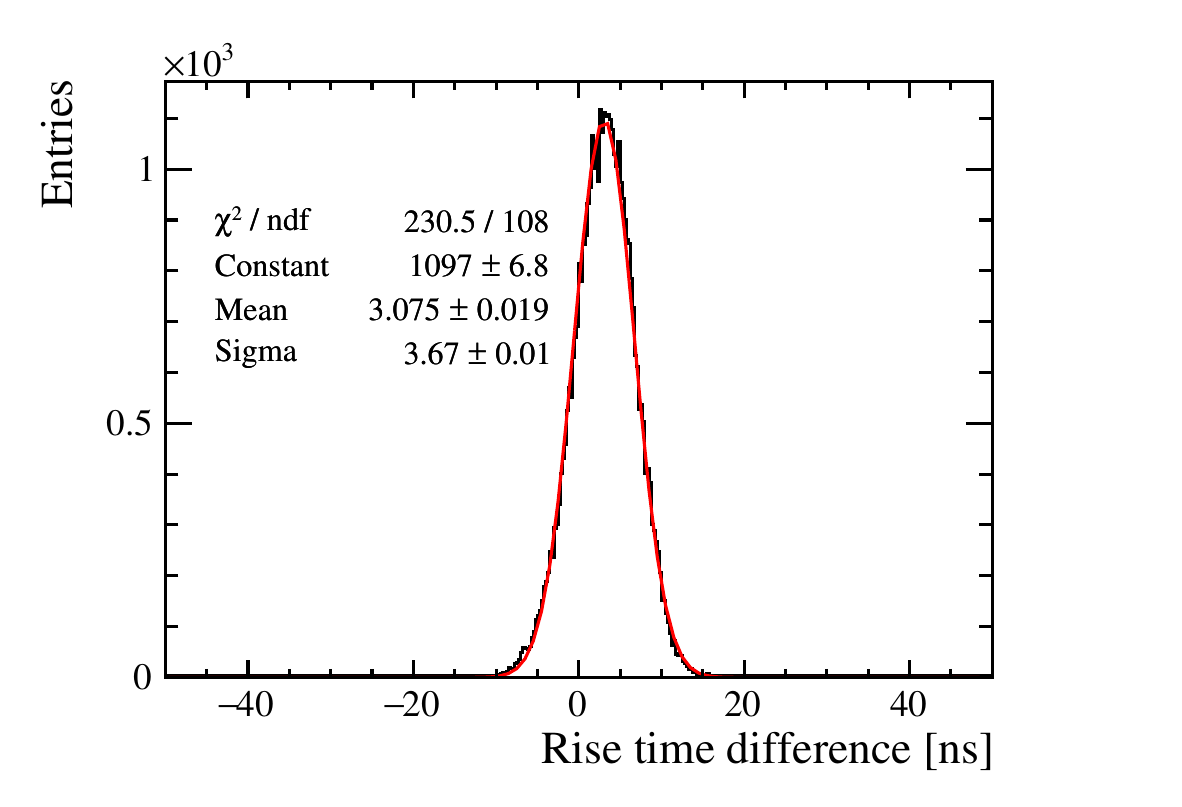}
        }
    \caption{First‑peak charge integration distribution of the time‑calibration data, where charges below 6000 ADC$\times$ns are regarded as dark noise. Difference in rise time between two channels, the distribution is Gaussian.}
    \label{fig:TCali_charge_diff}
\end{figure}

To comprehensively assess the long-term stability of the electronic system, we carried out 5 power-off restart operations and obtained 5 runs (Run1$\sim$5) of independent and valid operational data; each run was divided into 10 files, and each file contained 4096 trigger events. During the time calibration process, the signal arrival time of each channel was calculated based on the following conditions:

\begin{enumerate}
    \item The rise time is taken as the position where the waveform first reaches 20\% of the first peak height.
    \item Data with a first-peak integrated charge less than 6000 ADC$\cdot$ns are excluded to avoid dark noise, as shown in Figure~\ref{fig:TCali_charge_diff}.
\end{enumerate}

At the $m$-th flash of the LED, the photon arrival time ($T$) recorded by the $i$-th PMT can be expressed as:
\begin{equation}
	T_{mi}=t_{m}+TOF_{mi}+T_i+\tau_{mi}.
\end{equation}
Where $t_m$ denotes the time of the event relative to the starting point of signal recording, $TOF$ represents the photon time of flight within the detector, $T_i$ refers to the transit time and signal propagation time of the PMT, and $\tau$ encompasses the random fluctuations of each of the above parts, as well as the transit time spread (TTS) of the PMT and the instability of the LED's emission time.

Due to trigger settings, the statistical fluctuations of $t_m$ can be considerable. To minimize these statistical fluctuations, the difference between the signal times of two channels can be taken:
\begin{equation}
	T_{mi}-T_{mj}=TOF_{mi}-TOF_{mj}+T_{i}-T_{j}+\tau_{m,i}-\tau_{m,j}.
\end{equation}
Where, $T_i-T_j$ represents the time difference between PMT $i$ and $j$. For light emitted from the LED at the center of the detector, the time of flight for different PMTs is equal, and $\tau_{m,i}-\tau_{m,j}$ follows a Gaussian distribution centered at zero. Therefore, the parameter $T_{i}-T_{j}$ is defined as $\hat{\Delta T_{ij}}$, the mean value of the rise‑time difference distribution for any pair of channels, as shown in Figure~\ref{fig:TCali_charge_diff}. To prevent over-reliance on data from any single channel, we apply the following method to decentralize the time differences among all channels:
\begin{equation}
	\bar{T}_i=\frac{1}{N}\sum_{j=0}^N \hat{\Delta T_{ij}}.
\end{equation}
Where $N$ refers to the total living channel count, and $T_i$ is the time difference between channel $i$ and the mean time of all channels.
Because the position of the LED source is fixed while data acquisition, the statistical fluctuations contribute to the vast majority of the uncertainty of the difference of 2 independent runs, which is calculated as:
\begin{equation}
    \sigma_{T_{i,\alpha}-T_{i,\beta}}^2=\sigma_{T_{i,\alpha,stat}}^2+\sigma_{T_{i,\beta,stat}}^2.
\end{equation}
\begin{equation}
    \sigma_{T_{i,\xi,stat}}^2=\frac{1}{N^2}\sum_{j=0}^N \sigma_{\Delta T_{ij},\xi}^2, \forall \xi\in \left\lbrace\alpha,\beta\right\rbrace.
\end{equation}
Where $\alpha,\beta$ refer different data collection, the statistical uncertainty of $\Delta T_{ij}$ is obtained through the distribution width and statistics.

\begin{figure}[h]
    \centering
    \subfigure[Time offset]{
    \includegraphics[width=0.45\linewidth]{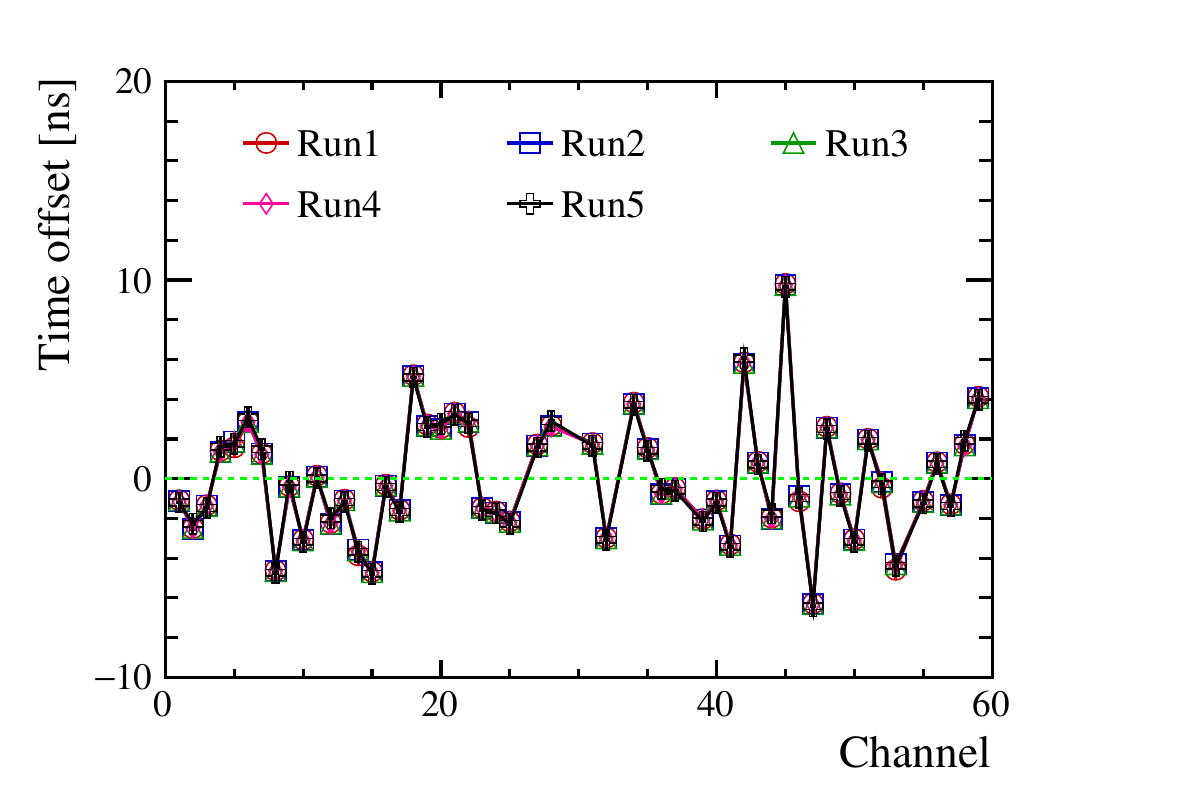}
    }
    \subfigure[Time offset difference]{
    \includegraphics[width=0.45\linewidth]{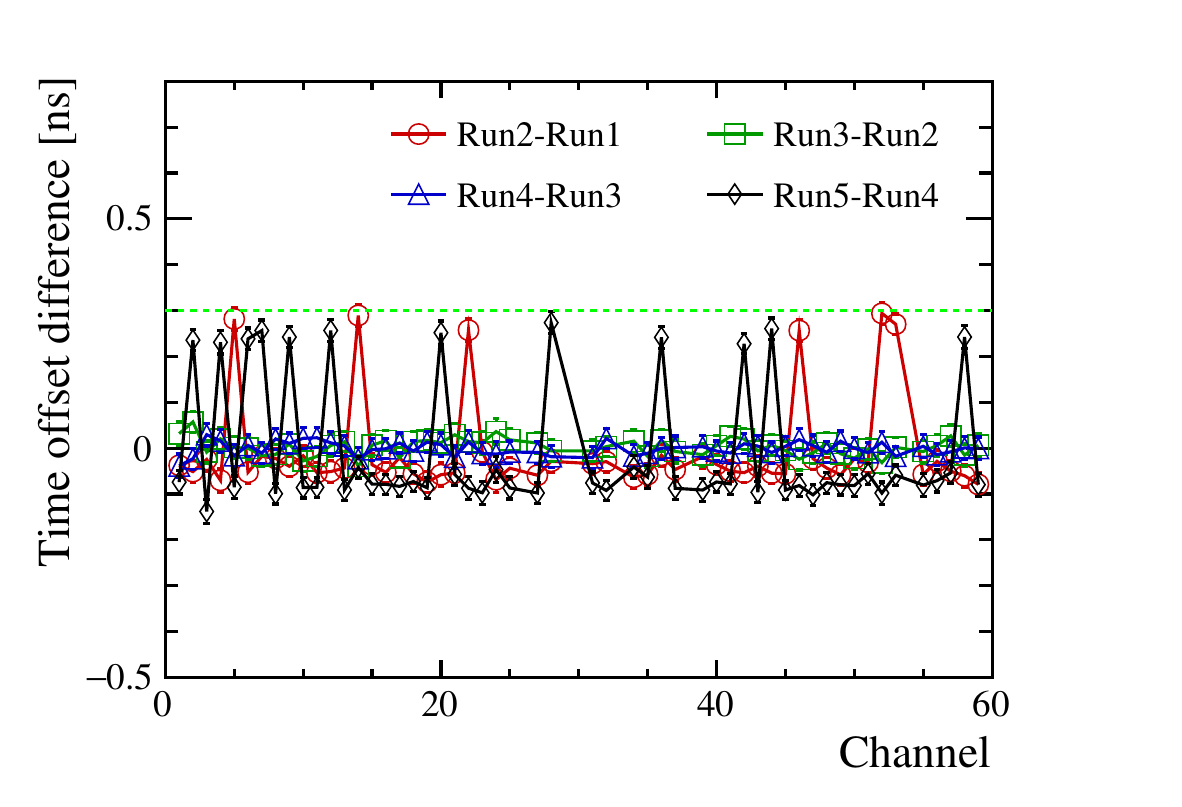}
    }
    \caption{The time calibration result of Run1$\sim$5 (left), and the difference in time calibration results for each channel among 5 Runs (right).  }
    \label{fig:clockshift}
\end{figure}

After independent time calibration of the collected data, a slight clock shift was observed, as shown in Figure~\ref{fig:clockshift}. Two instances of clock shift are found across the five runs. Specifically, one instance involved 16 channels on electronic boards, all exhibiting the same shift of $\sim$0.3 ns. This skew is attributed to the LMK04832 Phase-Locked Loop (PLL) operating in zero-delay mode. Upon power-up, the PLL automatically aligns the phase with a finite adjustment step of 166 ps, which accounts for the observed 0.3 ns clock skew.

\begin{figure}[h]
    \centering
    \includegraphics[width=0.49\linewidth]{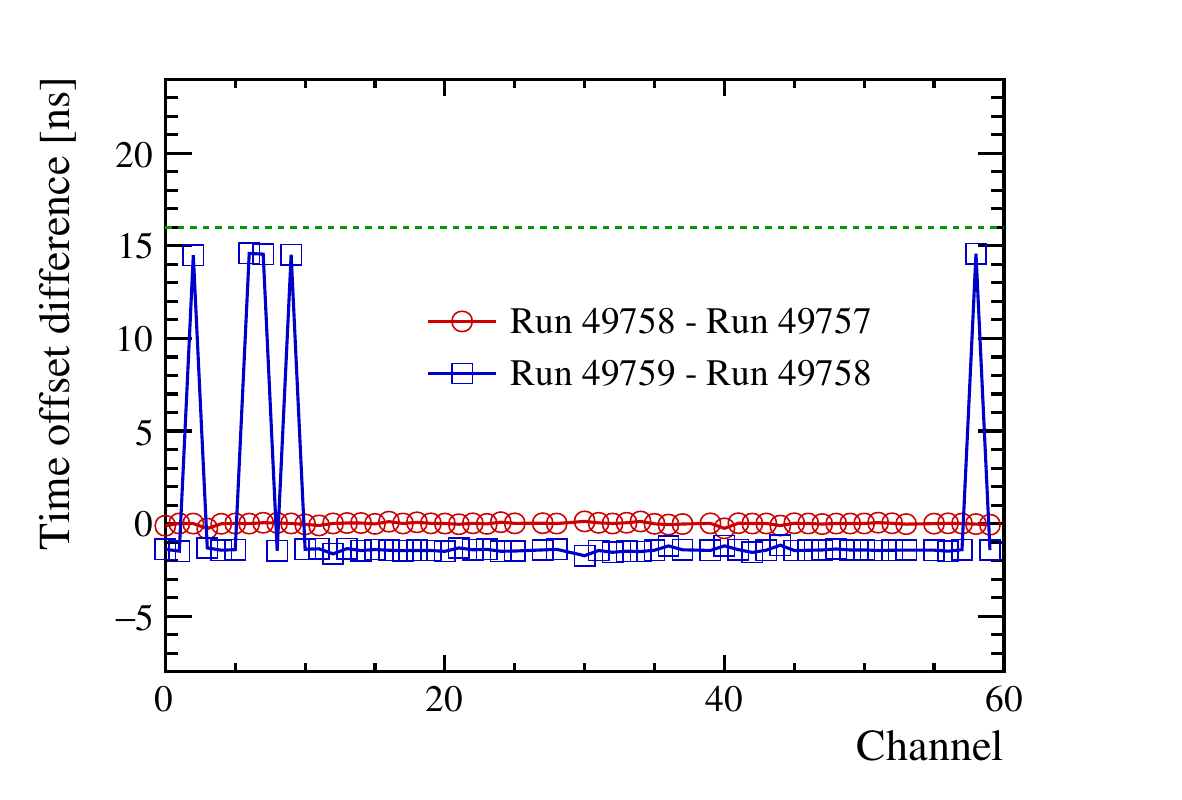}
    \caption{The difference of time calibration result for CAEN V1751 system. There is a 16 ns clock skew between run 49758 and run 49759. }
    \label{fig:clockshift_CAEN}
\end{figure}

By comparison, the same calibration method yielded a clock skew of 16 ns in the CAEN V1751 system, as shown in Figure~\ref{fig:clockshift_CAEN}, corresponding to its clock frequency of 125 kHz, which will significantly affect the event position reconstruction accuracy. The PDS1500 system has a time drift of only 0.3 ns—lower than the experimental time step (1 ns) and also lower than the PMT's transit time spread (TTS, 1.8 ns). 

\subsubsection{Energy threshold}
During normal data acquisition, limited by the relatively small target volume of the JNE-1ton detector, the signals originate mainly from natural radioactive backgrounds (U, Th, K) in the environment and detector materials, along with a small number of muon‑related signals.

The data transfer performance of the two systems and the design of their data acquisition programs determine the data acquisition rate. When using the CAEN V1751 system for data acquisition, limited by the data transfer bandwidth and the write rate of the acquisition program, the single‑channel trigger threshold was set to Tr$_{\text{single}}$= 5 CAEN ADC (approximately one single‑photoelectron signal), and the multi‑channel trigger threshold was set to Tr$_{\text{multi}}$= 45 (out of a total of 53 channels). Under these conditions, the event rate was approximately 280 Hz. 
When the multi‑channel threshold is continuously lowered, the acquisition program crashed, at which point the event rate was approximately 460 Hz.
\begin{figure}[h]
    \centering
    \includegraphics[width=0.75\linewidth]{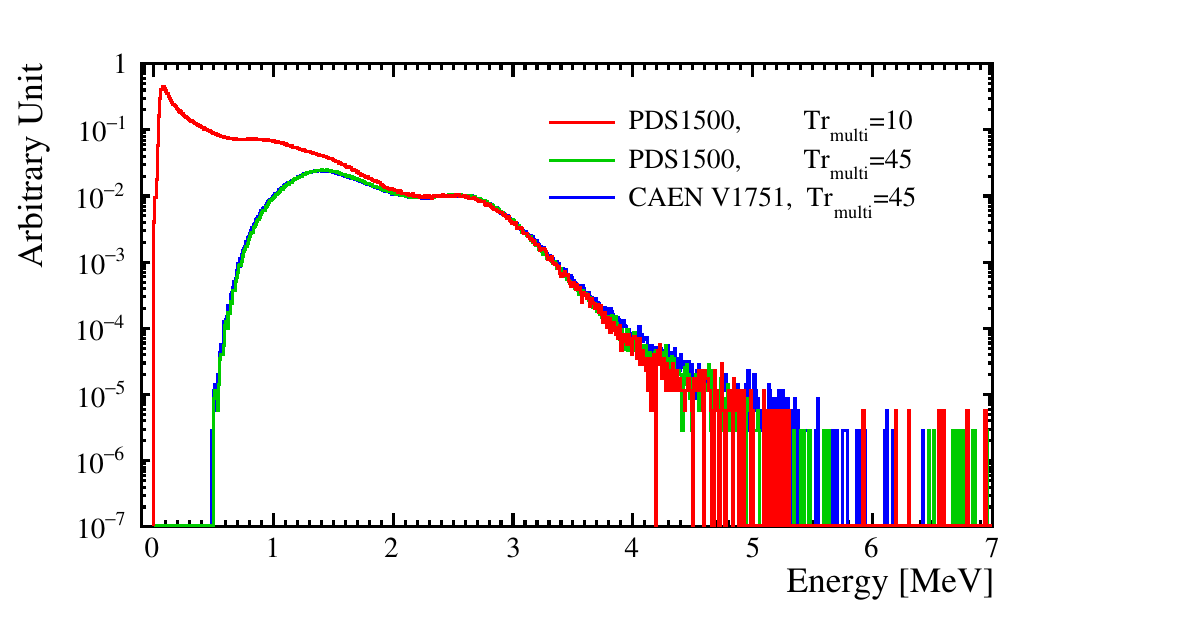}
    \caption{The reconstructed energy spectra obtained by analyzing different datasets with the same code~\cite{wu2023performance} and the upgrade detector setup itself, have their $^{208}$Tl (2.6 MeV) peaks aligned cross all spectra. 
    Among them, spectra with Tr$_{\text{multi}}$=45 are normalized spectra. For ease of comparison, the normalization method for spectrum with Tr$_{\text{multi}}$=10 is to keep the area within the [2.0, 3.0] MeV range consistent with that of Tr$_{\text{multi}}$=45 spectra.}
    \label{fig:energySpectrum}
\end{figure}

Under the combined constraints of the multi‑channel and single‑channel thresholds, the detection efficiency drops significantly below a certain energy, eventually approaching zero. Figure~\ref{fig:energySpectrum} clearly demonstrates the reduction in energy threshold achieved by the PDS1500 system by lowering the multi‑channel trigger threshold (Tr$_{\text{multi}}$=10): its detected spectrum only begins to drop significantly below about 0.1 MeV, whereas for other spectra with the trigger (Tr$_{\text{multi}}$=45) only start to appear above 0.5 MeV.

The lower energy threshold provides the detector with excellent detection performance, which is particularly advantageous for the measurement of low‑energy solar neutrinos. As shown in Figure~\ref{fig:energySpectrum}, when using the conventional CAEN V1751 system and its data acquisition program, the detection efficiency for energies below 1 MeV is significantly reduced, which would be detrimental to the neutrino measurement from the $pp$ chain and $^7$Be~\cite{WURM20171}.

\subsubsection{Energy resolution}
The energy resolution is one of the key parameters of a detector, and its quality directly affects the significance of physical observables. 
To precisely determine the energy resolution of the detector, an AmBe calibration source assembly is placed at the center of the detector during data acquisition to measure the neutron‑hydrogen capture signal. In this calibration assembly, the AmBe source is located at the center of a lead cube with a side length of 50 mm, and a small hole with a diameter of 8 mm is opened on the side of the lead cube. The 4.4 MeV gamma rays emitted by the AmBe source constitute the prompt signal; meanwhile, the emitted neutrons, after being moderated, are captured by hydrogen, releasing a 2.2 MeV gamma signal, which constitutes the delay signal.

The PDS1500 and CAEN V1751 systems independently acquired data from the detector with the AmBe calibration assembly placed inside. The thresholds of both systems are kept unchanged: the single‑channel trigger threshold is 5 CAEN ADC, and the multi‑channel trigger count is 45. During the actual data acquisition process, the data acquisition duration of the PDS1500 system is 3.1 hours, while that of the CAEN V1751 system is 19.3 hours.

For the acquired raw data, a similar code was used to analyze prompt‑delay coincidence signals~\cite{wu2023performance}. The screening condition set by the program is to pack all signals within a 2000 $\mu$s time window and extract signals with a pack count of 2 (double coincidence), containing one prompt signal and one delay signal, with the energy spectrum interval of the prompt signal being [3.8, 5.0] MeV. 

\begin{figure}[h]
    \centering
    \subfigure[PDS1500]{
    \includegraphics[width=0.45\linewidth]{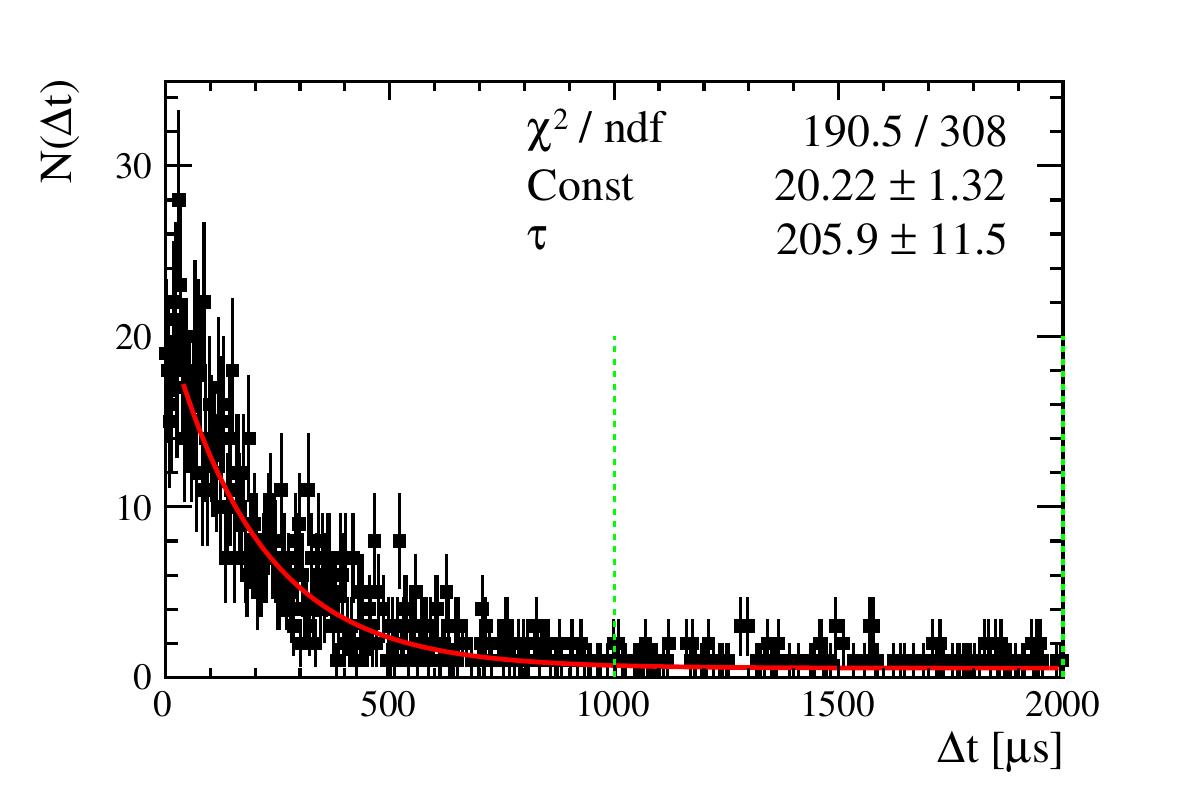}
    }
    \subfigure[CAEN V1751]{
    \includegraphics[width=0.45\linewidth]{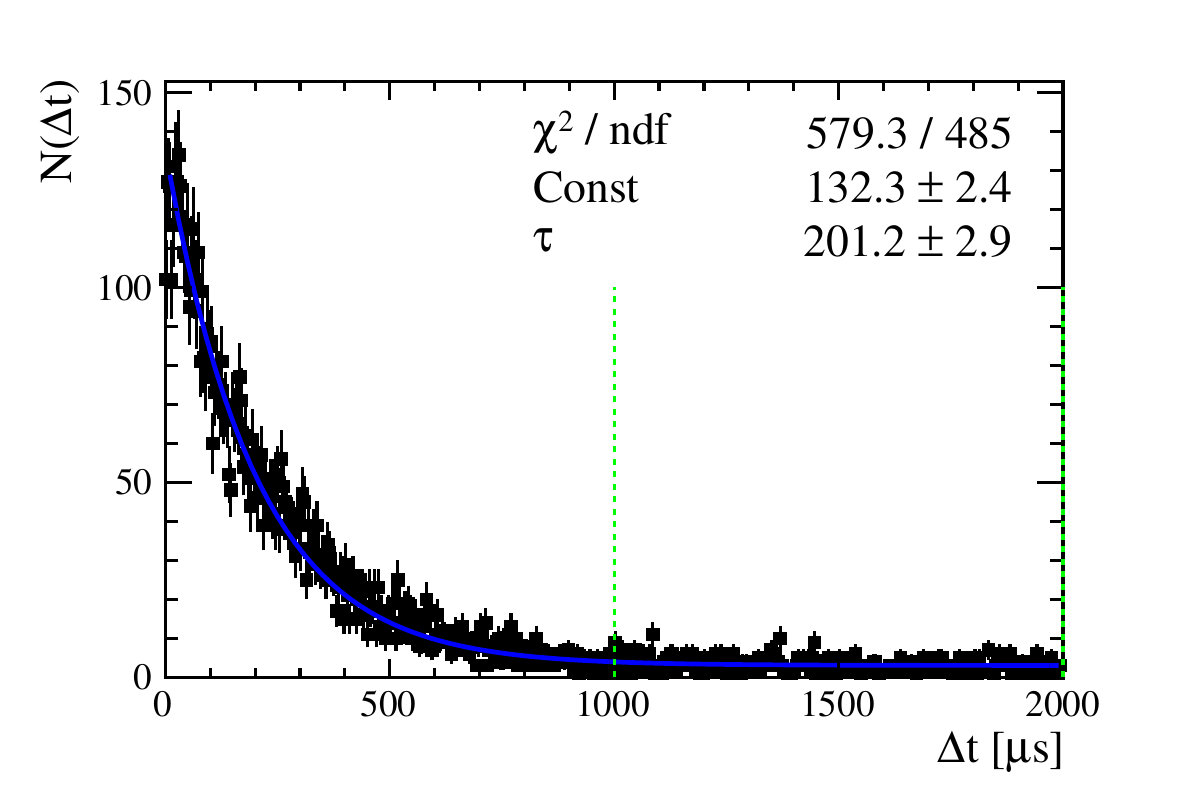}
    }
    \caption{The time difference ($\Delta t$) distributions of the prompt‑delay signals obtained using the similar packing code~\cite{wu2023performance} are shown. After fitting with an exponential plus constant function: $\text{Const}\times e^{-\frac{\Delta t}{\tau}}+\text{B}$, the neutron capture lifetime $\tau$ is obtained, the background $\text{B}$ is fixed as the mean count within the time window [1000, 2000] $\mu$s. }
    \label{fig:lifetime}
\end{figure}

Under these conditions, the time difference distributions of the prompt‑delay signals for the PDS1500 and CAEN V1751 systems are obtained, as shown in Figure~\ref{fig:lifetime}. The following formula is used to fit these distributions in the interval [30, 2000] $\mu$s:
\begin{equation}
	N(\Delta t)=\text{Const}\times e^{-\frac{\Delta t}{\tau}}+\text{B},
\end{equation}
where, $N(\Delta t)$ is the count corresponding to the time difference $\Delta t$
$\tau$ is the neutron capture time, and $\text{B}$ is the background count. The background $\text{B}$ is fixed as the mean count within the time window [1000, 2000] $\mu$s, their values are 0.540 and 3.056, respectively. The fitting results show that the neutron capture time $\tau$ is approximately 200 $\mu$s, which is consistent with theoretical expectations~\cite{DayaBay:2016ziq}.

\begin{figure}[h]
    \centering
    \subfigure[Neutron capture energy spectra]{
    \includegraphics[width=0.45\linewidth]{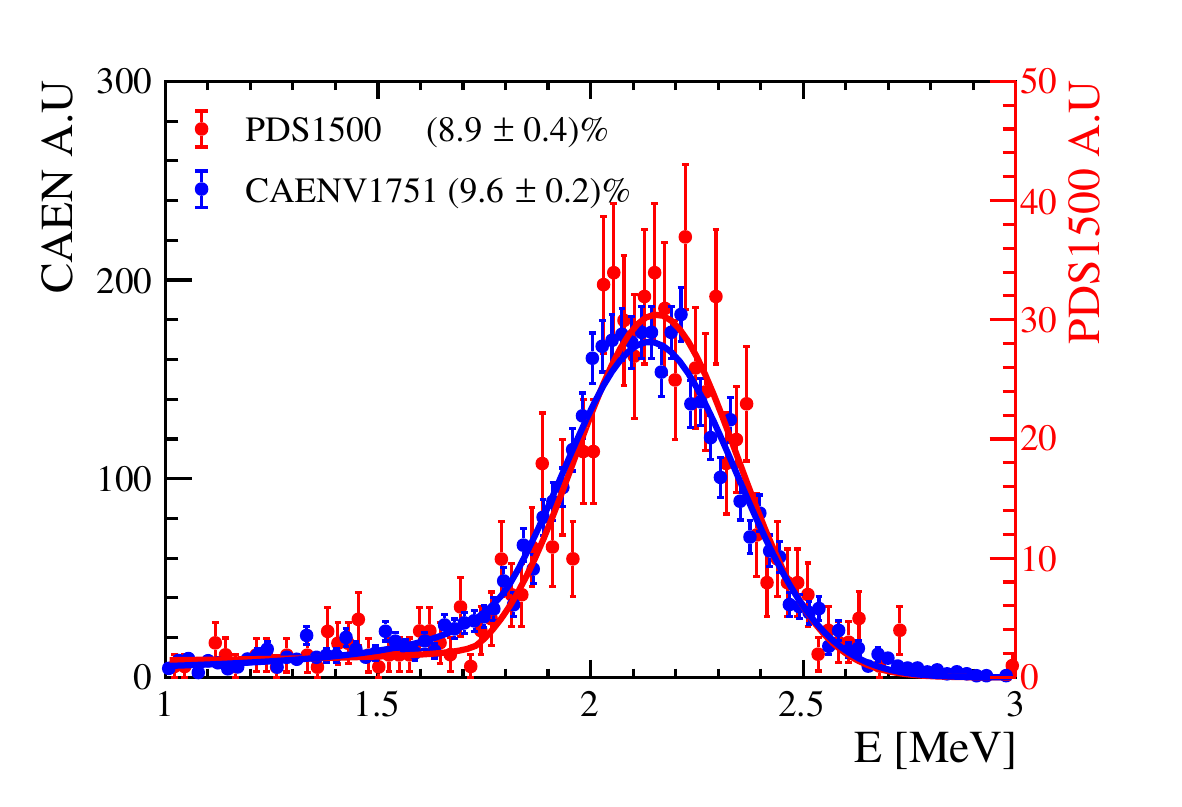}
    }
    \subfigure[Energy resolution vs bit number]{
    \includegraphics[width=0.45\linewidth]{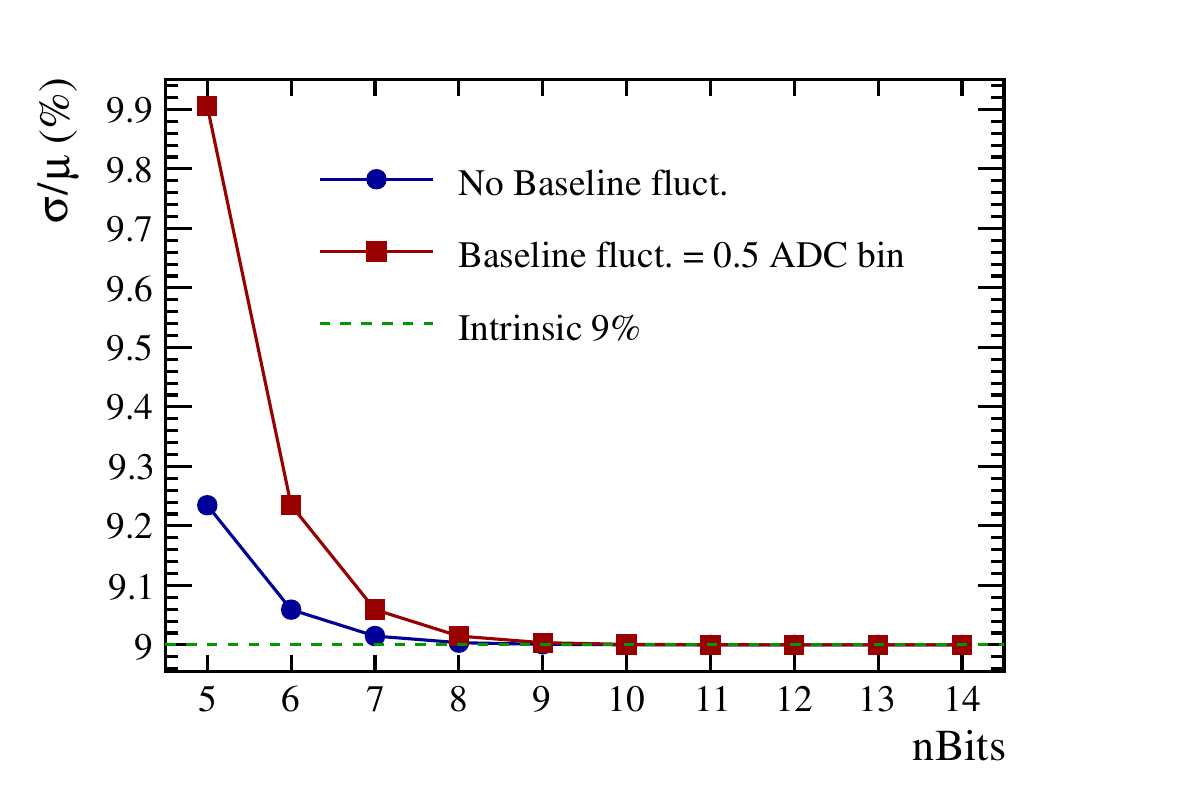}
    }
    \caption{(a)The neutron capture energy spectra measured by the PDS1500 and CAEN V1751 systems, which are fitted using the crystal ball function. (b)The energy resolution of the detector decreases with increasing ADC bit number of the electronics. When the bit number reaches 9 bits, the resolution approaches the limit at the intrinsic energy resolution 9.0\%. Meanwhile, the influence of baseline noise on the resolution is also significant. }
    \label{fig:resolution}
\end{figure}

By further selecting the delay signals, restricting them to within the fiducial volume ($R<$500 mm), time selection($\Delta t<200~\mu$s) and calculating and subtracting the accidental coincidence background, the energy spectrum of neutron capture on hydrogen was obtained, as shown in Figure~\ref{fig:resolution}(a). The crystal ball function \cite{DayaBay:2016ziq,gaiser_1982} fitting yields energy resolutions ($\sigma$/E) of (8.9 $\pm$ 0.4$_{stat}$)\% and (9.6 $\pm$ 0.2$_{stat}$)\% for the PDS1500 and CAEN V1751 systems, respectively.
The error arises solely from statistical fluctuations. Since the data are obtained from the same detector during the same operating period, and all conditions except for the electronics are essentially identical, the systematic error can be neglected in the comparative analysis.

Although the statistics of the PDS1500 system are lower than those of the CAEN V1751 system, the difference in the measured energy resolutions between the two systems is within 1.6 times the standard deviation; therefore, the energy resolutions of the two systems can be considered consistent.
Through a simple simulation as shown in Figure~\ref{fig:resolution}(b), we conclude that when the number of bits of the electronics exceeds 9, the energy resolution of the system stabilizes at around 9\% and does not improve further beyond this point. The intrinsic detector resolution is mainly determined by the light yield of the scintillator and the photocathode coverage of the PMT.

\section{JNE future}

The transition of the JNE detector from a 1‑ton prototype to a multi‑kiloton scale brings a dramatic increase in the number of PMTs, from 60 to 3,000. Such a leap in scale necessitates a commensurate evolution of the readout electronics, which must now satisfy much tighter performance requirements.
The future development of the electronics architecture focuses on addressing the exponential increase in channel density, the necessity for sophisticated data reduction, and the demand for sub-nanosecond synchronization across a distributed network. By leveraging next-generation FPGA technologies and high-speed communication protocols, the upgraded system aims to provide a robust, scalable, and high-throughput platform capable of capturing rare neutrino events with unprecedented precision. This section outlines the strategic roadmap for the JNE electronics, focusing on improvements in system integration, trigger intelligence, and timing distribution reliability.

\subsection{System Structure}
\begin{figure}[h]
\centering
\includegraphics[width=1.0\textwidth]{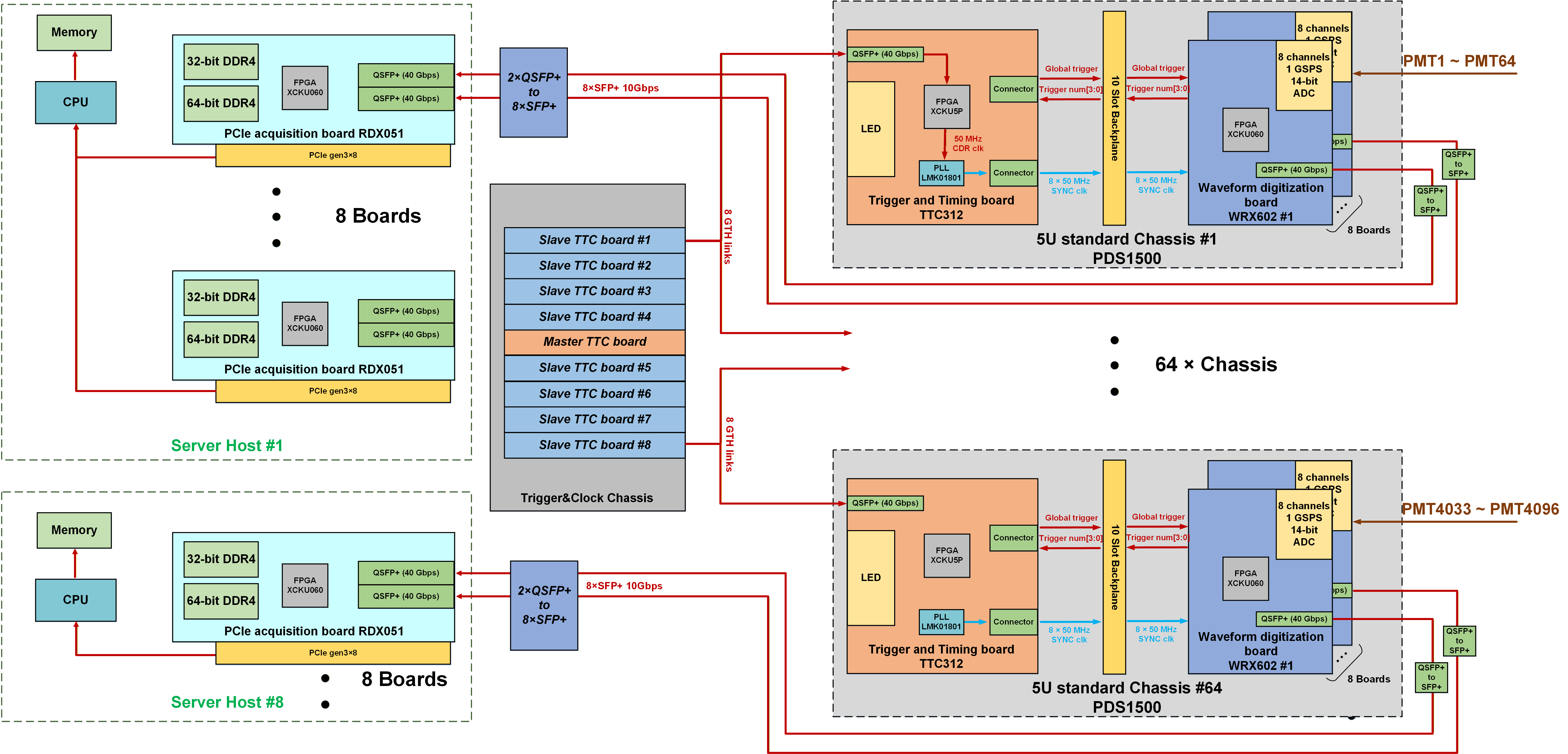}
\caption{The workflow of the 3000 channels electronics system.\label{fig:FutureArch}}
\end{figure}

To accommodate the requirements of the 3000-channel detector, the future electronics system will adopt a highly modular and hierarchical architecture. As illustrated in Figure~\ref{fig:FutureArch}, the system will be scaled to 64 standard 5U PDS1500 chassis, each housing eight WRX602 waveform digitization boards to process signals from PMTs. A centralized Trigger and Clock Chassis, comprising a master TTC board and eight slave TTC boards, serves as the synchronization core. This chassis will distribute synchronous clocks and manages global trigger logic across the entire detector through dedicated GTH links. 

Data aggregation will be performed through a two-stage optical fiber network. Waveform data from each digitization board will be first transmitted to a high-speed PCIe acquisition board (RDX051). 
These boards will be integrated into a high-performance server cluster consisting of eight server hosts, with each host supporting eight RDX051 boards. This structure not only ensures a massive total bandwidth of up to 2.56 Tbps for raw data throughput but also provides a flexible platform for real-time software processing and event building.

\subsection{Clock distribution structure}
\begin{figure}[h]
\centering
\includegraphics[width=0.49\textwidth]{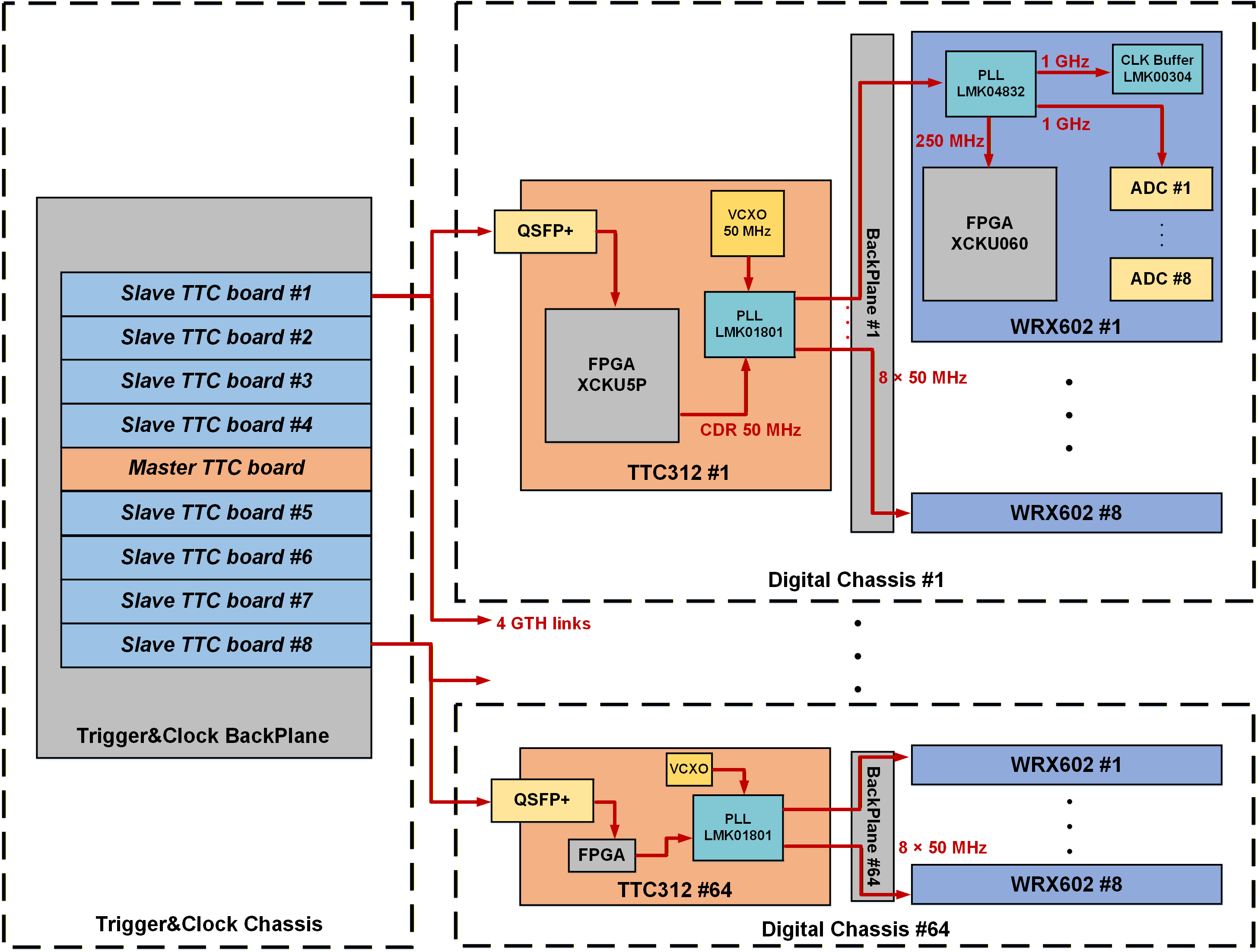}
\caption{The Clock distribution structure of the 3000 channels electronics system.\label{fig:Clock}}
\end{figure}

The precision of neutrino event reconstruction relies heavily on a unified and stable timing reference across all 3000 channels. As depicted in Figure~\ref{fig:Clock}, the future timing distribution will adopt a multi-stage tree topology centered on the Trigger and Clock Chassis. A Master TTC board will coordinate the synchronization, interfaced with eight Slave TTC boards via a dedicated backplane. The timing signals will be distributed to 64 digital chassis through high speed GTH optical links. At each digital chassis, the next-generation TTC312 board utilizes a high performance FPGA XCKU5P to perform Clock and Data Recovery (CDR), extracting a 50 MHz synchronous clock from the incoming serial data stream.

This recovered clock will be further processed by an onboard LMK01801 PLL and a 50 MHz VCXO to suppress jitter before being distributed to eight WRX602 digitization boards via the chassis backplane. Within each WRX602 board, an LMK04832 jitter cleaner performs the final frequency synthesis, generating a high-stability 1 GHz sampling clock for the ADCs and a 250 MHz reference clock for the FPGA logic. This hierarchical distribution scheme ensures that the sampling phases of all ADCs in the massive detector are tightly locked to the master reference, effectively minimizing the system-wide timing skew and ensuring high-resolution waveform digitization.

\subsection{Power consumption}
Regarding scalability, the PDS1500 prototype utilizes an eight-fan cooling configuration that effectively maintains core chip temperatures below 50 °C, demonstrating robust thermal management for a 160 W, 64-channel chassis. Scaling the system to 3000 channels would correspond to a total power consumption of $\sim$7.5 kW. While significant, this thermal load would be manageable within modern laboratory rack infrastructures. 

\section{Conclusion}

To meet the waveform acquisition and high-speed data transmission requirements of the future JNE experiment, which demands readout from approximately 3,000 PMTs, we have independently developed the PDS1500 electronics system—a single-crate 64-channel platform—and validated it on the upgraded JNE-1ton prototype detector against the commercial CAEN V1751 reference system.

Six key performance metrics were evaluated. The PDS1500 achieves zero data loss within the full 1000 ns acquisition window. Its baseline noise is reduced to one-third of the CAEN V1751 level, improving measurement stability. Timing calibration across five power-cycle restarts yields a clock drift of only 0.3 ns, compared with 16 ns in the CAEN V1751—a more than 50-fold improvement—well below both the 1 ns sampling step and the 1.8 ns PMT transit time spread. 
The energy detection threshold reaches as low as 0.1 MeV, versus $\sim$0.5 MeV for the CAEN V1751, enabling access to low-energy solar neutrinos from the pp chain and $^{7}$Be. Energy resolution of the neutron capture peak is consistent between the two systems at $\sim$9\%, a limit set by the intrinsic scintillator light yield and the photocathode coverage of PMT of the 1-ton prototype rather than electronics bit depth; higher physical resolution (14-bit) provides finer waveform detail but does not improve overall energy resolution in this configuration. 
The modular architecture provides the throughput and scalability required for the full 3,000-channel JNE detector. 

In summary, the PDS1500 fully satisfies all electronics requirements of the future JNE experiment.

\subsection* {Acknowledgments}
This work was supported in part by the National Natural Science Foundation of China (NSFC) under Grant 12127808 and 12505130, the Ministry of Science and Technology of China (No. 2022YFA1604704), and the China Postdoctoral Science Foundation (Certificate Number: 2024M751611).
\clearpage

\printcredits

\bibliographystyle{unsrt}

\bibliography{cas-refs}

\begin{thebibliography}{10}

\bibitem{Jinping:2016iiq}
John~F. Beacom et~al.
\newblock {Physics prospects of the Jinping neutrino experiment}.
\newblock {\em Chin. Phys. C}, 41(2):023002, 2017.

\bibitem{Ma:2021uzi}
Hao Ma et~al.
\newblock {Status and prospect of China Jinping Underground Laboratory}.
\newblock {\em J. Phys. Conf. Ser.}, 2156(1):012170, 2021.

\bibitem{Guo:2017nnr}
Ziyi Guo et~al.
\newblock {Slow Liquid Scintillator Candidates for MeV-scale Neutrino
  Experiments}.
\newblock {\em Astropart. Phys.}, 109:33--40, 2019.

\bibitem{Ouyang:2025phk}
Shuai Ouyang et~al.
\newblock {Development and characterization of the JNE concentrator}.
\newblock {\em Nucl. Instrum. Meth. A}, 1080:170755, 2025.

\bibitem{wu2023performance}
Yiyang Wu et~al.
\newblock {Performance of the 1-ton prototype neutrino detector at CJPL-I}.
\newblock {\em Nuclear Instruments and Methods in Physics Research Section A:
  Accelerators, Spectrometers, Detectors and Associated Equipment},
  1054:168400, 2023.

\bibitem{Luo:2023reconstruction}
Wentai Luo et~al.
\newblock {Reconstruction algorithm for a novel Cherenkov scintillation
  detector}.
\newblock {\em Journal of Instrumentation}, 18(02):P02004, 2023.

\bibitem{WANG2026170986}
Yuyi Wang et~al.
\newblock The fast stochastic matching pursuit for neutrino and dark matter
  experiments.
\newblock {\em Nuclear Instruments and Methods in Physics Research Section A:
  Accelerators, Spectrometers, Detectors and Associated Equipment},
  1082:170986, 2026.

\bibitem{SNO:2023cnz}
A.~Allega et~al.
\newblock {Event-by-event direction reconstruction of solar neutrinos in a high
  light-yield liquid scintillator}.
\newblock {\em Phys. Rev. D}, 109(7):072002, 2024.

\bibitem{BOREXINO:2021xzc}
M.~Agostini et~al.
\newblock {Correlated and integrated directionality for sub-MeV solar neutrinos
  in Borexino}.
\newblock {\em Phys. Rev. D}, 105(5):052002, 2022.

\bibitem{Xu:2022wcq}
Xun-Jie Xu et~al.
\newblock {Solar neutrino physics}.
\newblock {\em Prog. Part. Nucl. Phys.}, 131:104043, 2023.

\bibitem{Shao:2022yjc}
Wenhui Shao et~al.
\newblock {The potential to probe solar neutrino physics with LiCl water
  solution}.
\newblock {\em Eur. Phys. J. C}, 83(9):799, 2023.

\bibitem{Super-Kamiokande:2023jbt}
K.~Abe et~al.
\newblock {Solar neutrino measurements using the full data period of
  Super-Kamiokande-IV}.
\newblock {\em Phys. Rev. D}, 109(9):092001, 2024.

\bibitem{Bellini:2013wsa}
G.~Bellini et~al.
\newblock {Geo-neutrinos}.
\newblock {\em Prog. Part. Nucl. Phys.}, 73:1--34, 2013.

\bibitem{wan2017geoneutrinos}
Linyan Wan et~al.
\newblock Geoneutrinos at jinping: Flux prediction and oscillation analysis.
\newblock {\em Physical Review D}, 95(5):053001, 2017.

\bibitem{wang2020hunting}
Zhe Wang and Shaomin Chen.
\newblock Hunting potassium geoneutrinos with liquid scintillator cherenkov
  neutrino detectors.
\newblock {\em Chinese Physics C}, 44(3):033001, 2020.

\bibitem{DeGouvea:2020ang}
Andr\'e De~Gouv\^ea et~al.
\newblock {Fundamental physics with the diffuse supernova background
  neutrinos}.
\newblock {\em Phys. Rev. D}, 102:123012, 2020.

\bibitem{wei2017discovery}
Hanyu Wei et~al.
\newblock Discovery potential for supernova relic neutrinos with slow liquid
  scintillator detectors.
\newblock {\em Physics Letters B}, 769:255--261, 2017.

\bibitem{Dolinski:2019nrj}
Michelle~J. Dolinski et~al.
\newblock {Neutrinoless Double-Beta Decay: Status and Prospects}.
\newblock {\em Ann. Rev. Nucl. Part. Sci.}, 69:219--251, 2019.

\bibitem{fu2024comparison}
Hao-Yang Fu et~al.
\newblock Comparison study of counting and fitting methods in search for
  neutrinoless double beta decays.
\newblock {\em arXiv:2412.19859}, 2024.

\bibitem{AN2016133}
F.P. An et~al.
\newblock The detector system of the daya bay reactor neutrino experiment.
\newblock {\em Nuclear Instruments and Methods in Physics Research Section A:
  Accelerators, Spectrometers, Detectors and Associated Equipment},
  811:133--161, 2016.

\bibitem{BOGER2000172}
J~Boger et~al.
\newblock The sudbury neutrino observatory.
\newblock {\em Nuclear Instruments and Methods in Physics Research Section A:
  Accelerators, Spectrometers, Detectors and Associated Equipment},
  449(1):172--207, 2000.

\bibitem{Albanese_2021}
The~SNO+ collaboration et~al.
\newblock The sno+ experiment.
\newblock {\em Journal of Instrumentation}, 16(08):P08059, aug 2021.

\bibitem{ALIMONTI2009568}
G.~Alimonti et~al.
\newblock The borexino detector at the laboratori nazionali del gran sasso.
\newblock {\em Nuclear Instruments and Methods in Physics Research Section A:
  Accelerators, Spectrometers, Detectors and Associated Equipment},
  600(3):568--593, 2009.

\bibitem{Xilinx_KU060}
Xilinx.
\newblock {\em Kintex {UltraScale} {FPGAs} Data Sheet: {DC} and {AC} Switching
  Characteristics}, 2019.
\newblock DS892 (v1.19).

\bibitem{Xilinx_XC7Z020}
Xilinx.
\newblock {\em {Zynq-7000} {SoC} Data Sheet: Overview}, 2018.
\newblock DS190 (v1.11.1).

\bibitem{TI_LMK04832}
Texas Instruments.
\newblock {\em {LMK04832} Ultra Low-Noise {JESD204B} Compliant Clock Jitter
  Cleaner with Dual Loop {PLLs}}, March 2020.
\newblock Rev. E.

\bibitem{TI_LMK01801}
Texas Instruments.
\newblock {\em {LMK01801} Dual Clock Divider, Distributor, and Delay}, May
  2013.
\newblock Rev. C.

\bibitem{MeanWell_HRPG1000}
MEAN WELL.
\newblock {\em 1000W Single Output {AC/DC} Power Supply: {HRPG-1000} Series},
  2021.

\bibitem{Xilinx_KU5P}
Xilinx.
\newblock {\em Kintex {UltraScale+} {FPGAs} Data Sheet: {DC} and {AC} Switching
  Characteristics}, 2021.
\newblock DS922 (v1.18).

\bibitem{Zhang:2023ued}
Aiqiang Zhang et~al.
\newblock {Performance evaluation of the 8-inch MCP-PMT for Jinping Neutrino
  Experiment}.
\newblock {\em Nucl. Instrum. Meth. A}, 1055:168506, 2023.

\bibitem{PhysRevD.110.112017}
Xinshun Zhang et~al.
\newblock Study of neutron production for 360 gev cosmic muons.
\newblock {\em Phys. Rev. D}, 110:112017, Dec 2024.

\bibitem{Yang:2024yco}
Yuzi Yang.
\newblock {1-ton Prototype Neutrino Detector Upgrade at CJPL-I}.
\newblock {\em PoS}, ICHEP2024:227, 2025.

\bibitem{WURM20171}
Michael Wurm.
\newblock Solar neutrino spectroscopy.
\newblock {\em Physics Reports}, 685:1--52, 2017.
\newblock Solar Neutrino Spectroscopy.

\bibitem{DayaBay:2016ziq}
F.~P. An et~al.
\newblock {New measurement of $\theta_{13}$ via neutron capture on hydrogen at
  Daya Bay}.
\newblock {\em Phys. Rev. D}, 93(7):072011, 2016.

\bibitem{gaiser_1982}
J.~E. Gaiser.
\newblock {\em {Charmonium Spectroscopy from Radiative Decays of the $J/\psi$
  and $\psi'$}}.
\newblock PhD thesis, Stanford University, Stanford, CA, 1982.
\newblock SLAC-R-255.

\end{thebibliography}


\end{document}